\documentclass[11pt,a4paper]{article}
\pdfoutput=1

\usepackage{jcappub}
\usepackage[utf8]{inputenc}
\usepackage{graphicx}
\usepackage{verbatim}
\usepackage{epsfig}
\usepackage{subfigure}
\usepackage{color}
\usepackage{comment}
\usepackage{slashed}
\usepackage{amsmath}
\usepackage{ulem}

\DeclareMathOperator{\arcsinh}{arsinh}

\newcommand{\be}{\begin{equation}} 
\newcommand{\ee}{\end{equation}}
\newcommand{\bea}{\begin{eqnarray}} 
\newcommand{\eea}{\end{eqnarray}}
\newcommand{\nn}{\nonumber}

\newcommand{\bmp}{\noindent\begin{minipage}{16cm}}
\newcommand{\emp}{\end{minipage}\vskip 7mm} 
\def\lsim{\mathrel{\raise.3ex\hbox{$<$\kern-.75em\lower1ex\hbox{$\sim$}}}}
\def\gsim{\mathrel{\raise.3ex\hbox{$>$\kern-.75em\lower1ex\hbox{$\sim$}}}}

\newcommand{\intron}[1]{}

\title{Towards distinguishing variants of non-minimal inflation}

\author[a]{Tomo Takahashi}
\author[b,c]{and Tommi Tenkanen}

\affiliation[a]{Department of Physics, Saga University, Saga 840-8502, Japan}
\affiliation[b]{Department of Physics and Astronomy, Johns Hopkins University, \\
Baltimore, MD 21218, United States of America}
 \affiliation[c]{Astronomy Unit, Queen Mary University of London,
 \\ Mile End Road, London, E1 4NS, United Kingdom}

\emailAdd{tomot@cc.saga-u.ac.jp}
\emailAdd{ttenkan1@jhu.edu}

\abstract{We study models of inflation where the scalar field $\phi$ that drives inflation is coupled non-minimally to gravity via $\xi \phi^2 R$, or where the gravity sector is enlarged by an $R^2$ term. We consider the original Higgs inflation, Starobinsky inflation, and two different versions of a scenario where the inflaton is a scalar field other than the Higgs, and discuss if they can be distinguished from each other by measuring the tensor-to-scalar ratio and runnings of the spectral index of primordial curvature perturbations, on top of the amplitude and spectral index of the 
perturbations. We consider both metric and Palatini theories of gravity, showing how detailed studies of non-minimally coupled models can help to identify the inflaton field and how they may provide for a way to also distinguish between different theories of gravity in the present context.
}

\begin{document}

\maketitle


\section{Introduction}
\label{introduction}

Cosmic inflation is the current paradigm for explaining the origin of temperature fluctuations of the Cosmic Microwave Background radiation (CMB) and the large scale structure of the Universe~\cite{Starobinsky:1980te, Sato:1980yn, Guth:1980zm, Linde:1981mu, Albrecht:1982wi, Linde:1983gd}. An early period of accelerated expansion is also successful in explaining why the Universe is spatially flat, homogeneous, and isotropic to a high degree. Yet it remains unknown how inflation happened and how it ended, at least in our observable Hubble patch. 

Many phenomenological realizations of inflation have been studied in the literature over the last few decades~\cite{Lyth:1998xn,Mazumdar:2010sa,Martin:2013tda,Patrignani:2016xqp}, and among the parameters that are relevant to inflationary perturbations, two have already been measured to a high precision: the amplitude of the curvature power spectrum, $A_s=2.1\times 10^{-9}$, and the corresponding spectral tilt, $n_s\simeq 0.965$~\cite{Akrami:2018odb,Aghanim:2018eyx}. The Planck and BICEP2/Keck Array collaborations have also placed a strong constraint on the tensor-to-scalar ratio, $r<0.06$ at the reference scale $k_\ast = 0.05~{\rm Mpc}^{-1}$~\cite{Ade:2018gkx}.
In the future, this limit would become even more stringent.
For example, 
some on-going or near-future CMB B-mode polarization experiments
such as BICEP3 \cite{Wu:2016hul}, LiteBIRD~\cite{Matsumura:2013aja} and the Simons Observatory \cite{Simons_Observatory} are 
pushing the limit down to
$r\lesssim 0.001$ -- or aiming to detect $r$ above this limit. Yet other quantities such as the so-called running and running of the running of the spectral index $n_s$ have also been constrained by Planck~\cite{Akrami:2018odb}, although
currently the constraints are not severe. In the future, however, they are expected to improve by galaxy surveys and/or observations of the 21 cm line \cite{Kohri:2013mxa,Basse:2014qqa,Munoz:2016owz,Pourtsidou:2016ctq,Sekiguchi:2017cdy,Li:2018epc}.

Due to these good observational prospects, it is important to investigate in detail what we can learn about different inflationary models and how to distinguish between them. The data obtained in the past have already ruled out the most minimal models where inflation is driven by a scalar field (the `inflaton') slowly rolling in a simple power-law potential  (the so-called  chaotic inflation model), and in the future observations will allow to test also more complex models, for example those where the inflaton is coupled {\it non-minimally} to gravity or where the gravity sector is otherwise enlarged. This kind of models are arguably among the most interesting ones, as they are in very good agreement with the present data. Examples of such models include the famous Starobinsky~\cite{Starobinsky:1980te} and Higgs inflation~\cite{Salopek:1988qh,Bezrukov:2007ep} models. Especially the latter is a particularly interesting model, as it is based solely on the known Standard Model of particle physics (SM); for a recent review, see Ref. \cite{Rubio:2018ogq}.

However, as noted in many works in the past, it is not easy to distinguish between these models, as they both predict not only very small tensor-to-scalar ratio $r$ but also very similar numerical values for it. Furthermore, not only it is difficult to distinguish between Higgs and Starobinsky models but also between any other scalar field models which include a non-minimal coupling to gravity (see e.g. \cite{Spokoiny:1984bd,Futamase:1987ua,Fakir:1990eg,Amendola:1990nn,Kaiser:1994vs,Komatsu:1999mt,Bezrukov:2007ep,Bauer:2008zj,Park:2008hz,Lerner:2009xg,Linde:2011nh,Kaiser:2013sna,Kallosh:2013maa,Kallosh:2013tua,Chiba:2014sva,Boubekeur:2015xza,Pieroni:2015cma,Salvio:2017xul,Odintsov:2018qyy,Almeida:2018pir,Ferreira:2018nav}) -- sometimes regardless of the inflaton potential. This is because of the famous attractor behaviour \cite{Kaiser:2013sna,Kallosh:2013tua}, where independently of the original scalar potential some models asymptote to a universal attractor, the Starobinsky model, at the limit of large non-minimal coupling to gravity. Further complications in identifying the inflaton field arise because already within e.g. the Higgs inflation model, there can actually be more than one versions of it \cite{Bauer:2008zj,Germani:2010gm,Nakayama:2010kt,Kamada:2010qe,Kamada:2012se,Jinno:2017lun,Enckell:2018kkc,Rasanen:2018ihz}.

Some attempts to distinguish between different non-minimally coupled models of the same type have been made been in the past \cite{Lerner:2011ge,Bezrukov:2011gp} before the discovery of the Higgs boson. This is the goal of the present work, too. We will study four models: the original Higgs inflation, Starobinsky inflation, and two different versions of a scenario where the inflaton is a scalar field other than the Higgs, and will discuss how they can be distinguished from each other. In this paper we will augment the previous studies by adding three novel aspects which were not included in the previous works: first, we will study the inflationary dynamics also in the case where the inflaton field has a quadratic potential on top of its non-minimal coupling to gravity. Second, we will calculate the running and the running of the running of the spectral index of curvature perturbations, $\alpha_s, \beta_s$, and discuss to what extent the models can be distinguished from each other by observing $\alpha_s$ and/or $\beta_s$. Third, and most importantly, we will study inflationary dynamics in both {\it metric} and {\it Palatini} counterparts of gravity. 

In the usual metric formulation of gravity, one assumes that the space-time connection is determined by the metric only, i.e. it is the usual Levi-Civita connection. In the Palatini formalism, however, both the metric $g_{\mu\nu}$ and connection $\Gamma$ are treated as independent variables, on top of the matter degrees of freedom in the theory. In General Relativity (GR), the constraint equation for the connection imposes $\Gamma$ to be the Levi-Civita connection, and hence renders the two formalisms equivalent. However, with non-minimally coupled matter fields or otherwise enlarged gravity sector this is generally not the case~\cite{Sotiriou:2008rp}, and one has to make a choice regarding the underlying gravitational degrees of freedom\footnote{
As we will see, this does not necessarily amount to adding new degrees of freedom to the theory.
}. In the context of inflation, this will generically change the field dynamics and hence also the predictions of the model. This was originally noted in \cite{Bauer:2008zj}, and has recently gained increasing attention \cite{Bauer:2010jg,Tamanini:2010uq,Rasanen:2017ivk,Tenkanen:2017jih,Racioppi:2017spw,Markkanen:2017tun,Jarv:2017azx,Racioppi:2018zoy,Enckell:2018kkc,Carrilho:2018ffi,Enckell:2018hmo,Antoniadis:2018ywb,Rasanen:2018fom,Kannike:2018zwn,Rasanen:2018ihz,Almeida:2018oid,Antoniadis:2018yfq} (see also \cite{Azri:2017uor,Azri:2018gsz,Shimada:2018lnm}). Detailed studies of non-minimally coupled models are therefore interesting not only for distinguishing between different models of inflation, but also because they may provide for a way to distinguish between different theories of gravity. In this paper we will discuss to what extent this is possible in Higgs-like inflationary models.

The paper is organized as follows: in Section \ref{model}, we present the models under consideration and discuss the choice of gravitational degrees of freedom. In Sections \ref{inflation} and \ref{reheating}, we study 
inflationary dynamics and the subsequent reheating period in these models, paying particular attention to observables which might help in distinguishing between models of the same type. In particular, we will elaborate on the differences between the metric and Palatini counterparts of gravity, both at theoretical and observational level. In Section \ref{results}, we present the results of our numerical analysis and discuss the observational ramifications. Finally, in Section \ref{conclusions}, we conclude.


\section{The Models}
\label{model}

We will study extensions of the Standard Model where on top of the SM matter content we assume there is at least one scalar field, $S$, which can act as an inflaton field. The relevant part of the scalar potential then is
\be
\label{potential}
V(\Phi,S) = 
\lambda_{\rm H}(\Phi^\dagger\Phi)^2+\frac{m_{\rm S}^2}{2} S^2+\frac{\lambda_{\rm S}}{4}S^4
+ \frac{1}{2}\left(\xi_{\rm S}S^2 + \xi_{\rm H}\Phi^\dagger\Phi \right) g^{\mu\nu}R_{\mu\nu}(\Gamma) +\frac12 f(g_{\mu\nu},R_{\mu\nu}(\Gamma)),
\ee
where $\sqrt{2}\Phi^{\rm T}=(0,v+h)$ is the SM SU(2) gauge doublet in the unitary gauge and the Higgs mass has been omitted, as it does not play a role in inflationary dynamics. Likewise, as we will concentrate on cases where inflation is driven either by $h, S$ or the $f(g_{\mu\nu},R_{\mu\nu}(\Gamma))$ gravity term.
In this paper, we concentrate on the form of
\be
f(g_{\mu\nu},R_{\mu\nu}(\Gamma)) = f(R) = M_P^2 R + \alpha R^2,
\ee
which includes the so-called Starobinsky model. Here  $M_{\rm P}$ is the reduced Planck mass,
$g_{\mu\nu}$ is the metric tensor, $R_{\mu\nu}$ is the Ricci tensor, and $\Gamma$ is the connection. The Riemann tensor is constructed from the connection and its first derivatives,  $R^\rho_{\sigma\mu\nu}=R^\rho_{\sigma\mu\nu}(\Gamma,\partial\Gamma)$, and the Ricci tensor from this in the usual way, $R_{\mu\nu}=R^\lambda_{\mu\lambda\nu}$. 

The well-known but important and often unappreciated point is that no metric is needed to construct geometry. Usually one simply {\it assumes} that the underlying theory of gravity is of {\it metric} type, so that the connection is given by
\be
	\bar{\Gamma}^\lambda_{\alpha\beta} = \frac{1}{2}g^{\lambda\rho}(\partial_\alpha g_{\beta\rho} + \partial_\beta g_{\rho\alpha} - \partial_\rho g_{\alpha\beta}) \,.
\ee
This is the Levi-Civita connection, which associates the metric uniquely with the connection. In the so-called {\it Palatini} formalism, however, both $g_{\mu\nu}$ and $\Gamma$ are treated as independent variables\footnote{
In the following we will assume, for simplicity, that the connection is torsion-free, $\Gamma^\lambda_{\alpha\beta}=\Gamma^\lambda_{\beta\alpha}$. For non-vanishing torsion, see e.g. Refs. \cite{Rasanen:2018ihz,Shimada:2018lnm}, where it was shown that the new terms the non-zero torsion introduces can significantly complicate the analysis and may change the overall results. The effect of that is, however, beyond the scope of this work.}. In GR,  the constraint equation for the connection (which is found by varying the action with respect to the connection) imposes $\Gamma$ to be the Levi-Civita connection and hence renders the two formalisms equivalent. However, with non-minimally coupled matter fields or otherwise enlarged gravity sector this is generally not the case~\cite{Sotiriou:2008rp}. For example, in the case of a non-minimally coupled matter field $\phi$, variation of the action with respect to the connection gives \cite{Bauer:2008zj}
\be
\label{connection_solution}
\Gamma = \bar{\Gamma} + \delta^\lambda_\alpha\partial_\beta\omega(\phi) + \delta^\lambda_\beta\partial_\alpha\omega(\phi) - g_{\alpha\beta}\partial^\lambda\omega(\phi) ,
\ee
where $\omega(\phi)=\ln\sqrt{1+\xi\phi^2/M_{\rm P}^2}$. From Eq. \eqref{connection_solution}, deviation from the Levi-Civita connection is clear. Therefore, it is a natural starting point to consider a theory where geometry depends on both the metric and the matter fields coupled non-minimally to gravity. One can even argue that the Palatini models we consider in this paper should not be seen as examples of `modified gravity' theories, as currently we do not know what the fundamental gravitational degrees of freedom are. 

In any case, as we will discuss in detail in the next section, this means that models of cosmic inflation in the Palatini and metric formulations are intrinsically different, with profound consequences on inflationary observables. However, as can be seen from the form of Eq.~\eqref{connection_solution}, when the inflaton field relaxes down to a value $\phi\ll M_{\rm P}/\sqrt{\xi}$ after inflation, we retain the pure GR form of the theory. Hence, the model will pass all precision tests of gravity in the present universe. By studying in detail the inflationary dynamics of such models, however, one may be able to distinguish not only between different models of inflation but also different formulations of gravity, despite them giving the same theory at late times. This is exactly what we aim to do in this paper.

When it comes to the assumed form of the non-minimal couplings between scalar fields and gravity, they are motivated by the analysis of quantum corrections in a curved background. It has been shown that renormalizability requires terms of the above form even if the coupling constants $\xi_i, i=H,S$ are initially set to zero \cite{Birrell:1982ix}. This is the reason for considering in this paper only such non-minimal couplings which involve operators quadratic in the field. There are some mild observational constraints on the magnitudes of such couplings: in the case of the SM Higgs, LHC constraints require $\xi_{\rm H} \lesssim 10^{15}$ in the metric case \cite{Atkins:2012yn}. As we will show, however, in the Palatini case such constraint cannot be derived. For $\xi_{\rm S}$, no observational constraints exist beyond the requirement that in the present Universe $\xi_{\rm S} \langle S^2\rangle \ll M_{\rm P}^2$.

Before discussing the models in more detail, we comment on the choice of the gravitational degrees of freedom. Despite the fact that the Palatini versions of $f(R)$ theories suffer from certain theoretical and observational shortcomings (see e.g. \cite{LanahanTremblay:2007sg,Sotiriou:2008rp}), 
Palatini models where a non-minimally coupled scalar field has a canonical kinetic term do not constitute an $f(R)$ theory, and are therefore free of the problems that Palatini counterparts of $f(R)$ theories may face.  Instead, such models can be shown to constitute a `degenerated higher order scalar-tensor theory' of a certain type \cite{Aoki:2018lwx} or a `non-trivial Brans-Dicke theory' \cite{Capozziello:2015lza}, which in our case are equivalent to a metric theory with a non-canonical kinetic term for the scalar field \cite{Koivisto:2005yc}. Because of that, this type of models are also sometimes called `metric-affine' theories \cite{Azri:2017uor,Azri:2018gsz,Rasanen:2018ihz,Shimada:2018lnm}. However, as discussed above, non-minimal couplings to gravity should be seen not as an {\it ad-hoc} addition to inflationary models but as a generic ingredient of quantum field theory (QFT), either because they are generated radiatively or because no symmetry arguments forbid including them in the first place. As in usual QFT the matter fields have {\it canonical} kinetic terms in the Jordan frame, it is in this sense that one can say that the differences between the cases which we will call `metric' and `Palatini' are indeed in the underlying theory of gravity and not simply in the choice of the model, that is in the choice of the scalar field kinetic term and potential. While at least at classical level one could simply start in the Einstein frame with any unusual scalar potential and/or kinetic term introduced by hand, it is the surprising connection with gravity that makes only certain models particularly interesting. In this paper, we will discuss a subset of such models.

Finally, we present the models. The different models we will study are:
\begin{enumerate}
\item Higgs inflation ($\lambda_{\rm H}(\Phi^\dagger\Phi)^2$ dominates; $m_{\rm S}, \lambda_{\rm S}, \alpha = 0$),
\item Starobinsky inflation ($f(g_{\mu\nu},R_{\mu\nu}(\Gamma))$ dominates; $m_{\rm S},\lambda_{\rm S} =0; h\simeq 0$),
\item Quadratic S-inflation ($m^2_{\rm S}S^2$ dominates; $\lambda_{\rm S}, \alpha = 0; h\simeq 0$),
\item Quartic S-inflation ($\lambda_{\rm S}S^4$ dominates; $m_{\rm S}, \alpha =0; h\simeq 0$) ,
\end{enumerate}
where the expressions in parenthesis denote the corresponding choices of parameters\footnote{
As the SM Higgs quartic self-coupling $\lambda_{\rm H}$ does not vanish in general, we have in cases other than the Higgs inflation assumed that instead of imposing $\lambda_{\rm H}=0$, the field itself is non-dynamical during inflation, $h\simeq 0$.
}. The first model, Higgs inflation, has originally been discussed in Refs. \cite{Bezrukov:2007ep} (metric case) and \cite{Bauer:2008zj} (Palatini case), the Starobinsky inflation\footnote{
As we will show, Starobinsky inflation does not have a Palatini counterpart.
} with $f(g_{\mu\nu},R_{\mu\nu}(\Gamma))\propto R^2$ in \cite{Starobinsky:1980te}, and non-minimal inflation with a $Z_2$ symmetric polynomial potential in \cite{Futamase:1987ua,Salopek:1988qh} (metric) and \cite{Bauer:2008zj,Tenkanen:2017jih,Jarv:2017azx} (Palatini). In this paper we will study only the `pure' cases shown above; mixed models where scalar fields couple non-minimally to an extended gravity sector have been studied recently in e.g. \cite{Salvio:2015kka,Calmet:2016fsr,Wang:2017fuy,Ema:2017rqn,He:2018gyf,Ghilencea:2018rqg,Enckell:2018hmo,Antoniadis:2018ywb,Gundhi:2018wyz,Karam:2018mft,Antoniadis:2018yfq,Enckell:2018uic}, and multifield models with non-minimal couplings to gravity in e.g. \cite{Kaiser:2013sna,Schutz:2013fua,Kallosh:2013daa,Carrilho:2018ffi,Almeida:2018oid}.


\section{Inflation}
\label{inflation}

We begin by comparing a generic model of cosmic inflation with a non-minimal coupling to gravity in the metric and Palatini cases. We will first concentrate on Higgs-like models in Section \ref{higgslike}, and study Starobinsky inflation in Section \ref{starobinsky}.


\subsection{Higgs-like inflation}
\label{higgslike}

Assuming that only one field is dynamical during inflation, the action reads
\be \label{nonminimal_action1}
	S_J = \int d^4x \sqrt{-g}\left(\frac{1}{2}\left(M_{\rm P}^2 + \xi \phi^2\right) g^{\mu\nu}R_{\mu\nu}(\Gamma) + \frac{1}{2} g^{\mu\nu}\partial_{\mu}\phi\partial_{\nu}\phi - V(\phi) \right) \,,
\ee
where 
$g$ is the determinant of the metric tensor, and $\phi=h$ or $\phi=S$. The non-minimal coupling in the Jordan frame action \eqref{nonminimal_action1} can be removed by a Weyl transformation
\be \label{Omega1}
	g_{\mu\nu} \to \Omega(\phi)^{2}g_{\mu\nu}, \hspace{1cm} \Omega^2(\phi)\equiv 1+\frac{\xi \phi^2}{M_{\rm P}^2} \,,
\ee
which gives
\be \label{einsteinframe1}
	S_E = \int d^4x \sqrt{-g}\left(\frac{1}{2}M_{\rm P}^2g^{\mu\nu}R_{\mu\nu}(\Gamma)-\frac{M_{\rm P}^2+\xi \phi^2+
	6\kappa \xi^2\phi^2}{2\Omega^2(\phi)\left(M_{\rm P}^2+\xi\phi^2\right)}\, g^{\mu\nu} \partial_{\mu}\phi\partial_{\nu}\phi - \frac{V(\phi)}{\Omega(\phi)^4} \right) \,,
\ee
where $\kappa=1$ in the metric case and $\kappa=0$ in the Palatini case. Because in this frame the non-minimal coupling to gravity vanishes, this frame is called the Einstein frame. Because now the connection appears only in the Einstein-Hilbert term, in this frame $\Gamma = \bar{\Gamma}$, i.e. one retains the Levi-Civita connection. Note that Eq. \eqref{einsteinframe1} exhibits the fact that Palatini models are equivalent to metric theories with a non-canonical kinetic term for the scalar field in the Einstein frame, as discussed in Sec. \ref{model}. With zero torsion, the only difference between the two theories of gravity is therefore in the value of $\kappa$. However, as also discussed in Sec. \ref{model}, while one could, in principle, start in the Einstein frame with any kind of kinetic term, we retain the usual assumption that the kinetic terms are always canonical in the frame where the theory is defined.

With a suitable field redefinition $\phi = \phi(\chi)$, determined by 
\be \label{chi1}
	\frac{d\phi}{d\chi} = \sqrt{\frac{\Omega^2(\phi)\left(M_{\rm P}^2+\xi \phi^2\right)}{M_{\rm P}^2+\xi \phi^2+6\kappa \xi^2\phi^2}} \,,
	\ee
the Einstein frame kinetic term can be brought back to a canonical form\footnote{
Strictly speaking this applies only to the one-field case; for multi-field cases, see e.g. \cite{Lerner:2009xg,Carrilho:2018ffi,Almeida:2018oid}. The form of the field redefinition also shows that in the Palatini case where $\kappa=0$, no collider constraints on $\xi_{\rm H}$  similar to those derived in \cite{Atkins:2012yn} can be obtained. This is due to the fact that in \cite{Atkins:2012yn} the authors considered only the metric case and derived limits for the quantity $\beta=6\xi_{\rm H}^2v^2/M_{\rm P}^2$, which is quadratic in $\xi_{\rm H}$. As is evident from Eq. \eqref{chi1}, this quantity only appears in the metric case, and in Palatini gravity one can only constrain quantities like $\xi_{\rm H}v^2/M_{\rm P}^2$, for which one obtains roughly $\xi_{\rm H} \lesssim \mathcal{O}(10^{32})$ -- not a particularly stringent constraint.
}. The solution to this is \cite{GarciaBellido:2008ab,Bauer:2008zj,Rasanen:2017ivk}
\be
\label{chi_solution}
\frac{\sqrt{\xi}}{M_{\rm P}}\chi = \sqrt{1+6\kappa\xi}\sinh^{-1}\left(\sqrt{1+6\kappa\xi}u\right) - \sqrt{6\xi}\kappa\sinh^{-1}\left(\sqrt{6\xi}\frac{u}{\sqrt{1+u^2}}\right) ,
\ee
where $u\equiv \sqrt{\xi}\phi/M_{\rm P}$.
The action~\eqref{einsteinframe1} then becomes
\be \label{EframeS1}
	S_{\rm E} = \int d^4x \sqrt{-g}\bigg(\frac{1}{2}M_{\rm P}^2R -\frac{1}{2}{\partial}_{\mu}\chi{\partial}^{\mu}\chi - U(\chi)  \bigg) \,,
\ee
where $U(\chi) =V(\phi(\chi))/ \Omega^{4}(\phi(\chi))$ and $R = g^{\mu\nu}R_{\mu\nu}(\bar{\Gamma})$.  In this model inflation takes place at large field values for which the Einstein frame potential either develops a plateau (in the case of quartic potential in the Jordan frame) or has an otherwise suitable flattening effect (in the case of quadratic potential). One can also see that when $\chi\to 0$, the usual Einstein-Hilbert gravity of GR is retained regardless of the choice of formalism (metric or Palatini).

\begin{figure}
\begin{center}
\includegraphics[width=.48\textwidth]{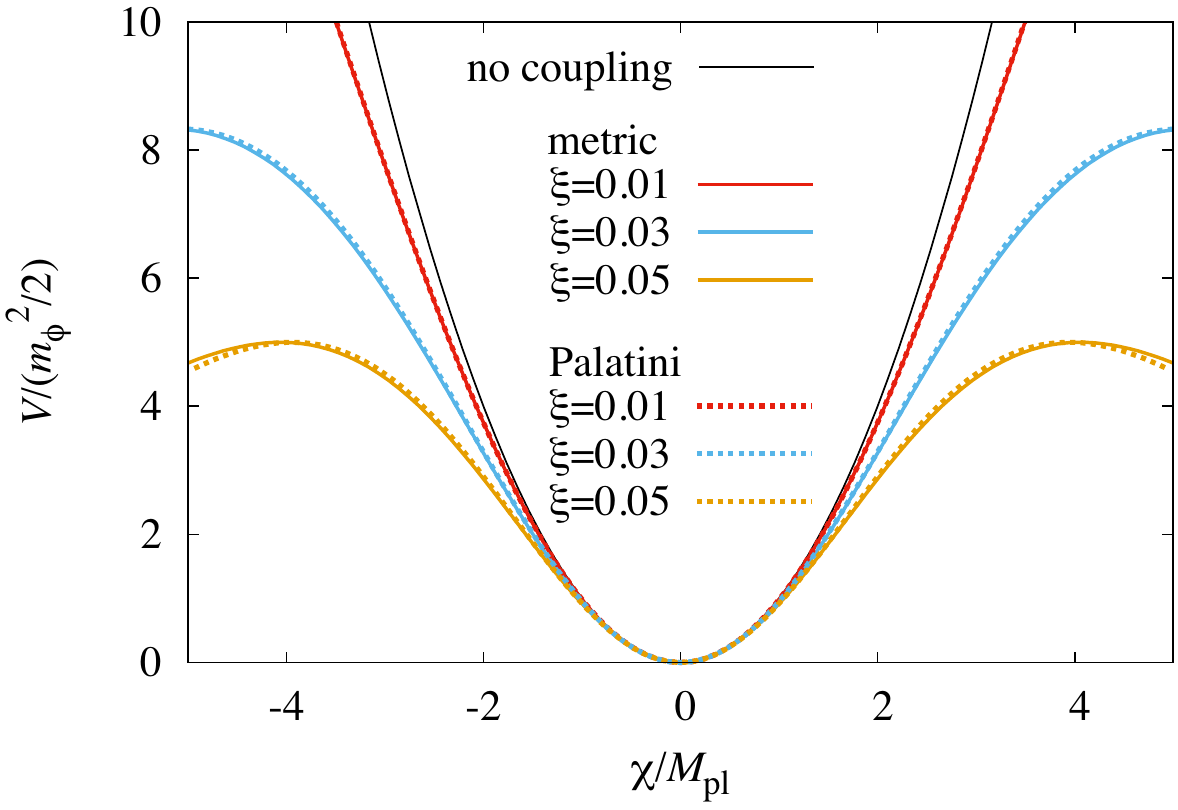}\hspace{5mm}
\includegraphics[width=.48\textwidth]{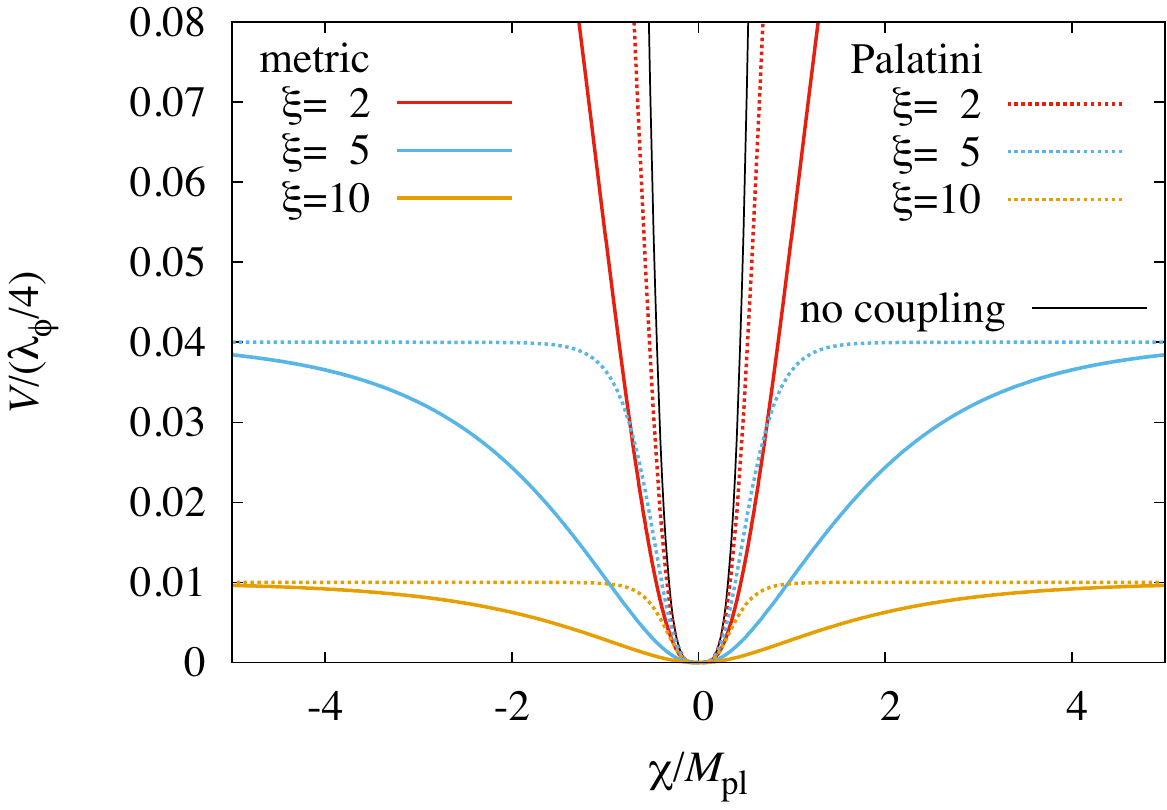}
\caption{Einstein frame potentials for the cases with the Jordan frame potential  assumed as $V(\phi) = (m^2 /2)\phi^2$ (left) and $V(\phi) = (\lambda/ 4) \phi^4$ (right). 
Those for the metric and Palatini cases are shown for several values of $\xi$. Notice that the metric and the Palatini cases are almost identical for the quadratic potential.}
\label{potential_plots}
\end{center}
\end{figure}

Let us first consider the cases where the Jordan frame potential is quartic in the field~$\phi$, i.e. $V(\phi)=\lambda \phi^4/4$.  In that case the canonically normalized field can be expressed as
\bea
\phi(\chi) 
\begin{cases} 
\simeq \displaystyle\frac{M_{\rm P}}{\sqrt{\xi}} \exp\left(\sqrt{\frac{1}{6}}\frac{\chi}{M_{\rm P}} \right)	  & \quad \mathrm{metric} ,\\   
= \displaystyle\frac{M_{\rm P}}{\sqrt{\xi}}\sinh\left(\frac{\sqrt{\xi}\chi}{M_{\rm P}}\right)& \quad \mathrm{Palatini} , 
\end{cases}
\eea
and hence the large field Einstein frame potential reads
\bea \label{chipotential1}
	U^{(4)}(\chi) =\frac{\lambda}{4} \frac{\phi^4(\chi)}{\Omega^4(\phi(\chi))} 
\begin{cases}	
\simeq \displaystyle\frac{\lambda M_{\rm P}^4}{4\xi^2}
\bigg(1+\exp\left(-\sqrt{\frac{2}{3}} \displaystyle\frac{\chi}{M_{\rm P}} \right) \bigg)^{-2}
	& \quad \mathrm{metric} ,\\ 
	= \displaystyle\frac{\lambda M_{\rm P}^4}{4\xi^2}
	\tanh^4\left(\displaystyle\frac{\sqrt{\xi}\chi}{M_{\rm P}}\right)& \quad \mathrm{Palatini} , 
\end{cases}
\eea
where the expressions in the metric case apply for $\xi\gg 1$ and $\chi\gg \sqrt{3/2}M_{\rm P}$, whereas the expressions in the Palatini case are exact. However, in our numerical analysis we do not use the approximate result \eqref{chipotential1} but compute everything using the exact result $U(\chi) =V(\phi(\chi))/ \Omega^{4}(\phi(\chi))$ with $\Omega(\phi)$ given by Eq.~\eqref{Omega1} and $\phi(\chi)$ by 
Eq.~\eqref{chi1}. 
In Fig.~\ref{potential_plots}, we show how the Einstein frame potential changes depending on the value of $\xi$ for the metric and Palatini cases. In the left and right panels, 
the cases with the Jordan frame potential $V(\phi) = (m^2/2) \phi^2$ and $ V(\phi) = (\lambda/4) \phi^4$ are depicted.
From the result \eqref{chipotential1} we see that for $\chi\gg \sqrt{3/2}M_{\rm P}$ in the metric case or $\chi\gg M_{\rm P}/\sqrt{\xi}$ in the Palatini case, the potential tends to a constant exponentially fast and is therefore suitable for slow-roll inflation (see the right panel of Fig.~\ref{potential_plots}). As we will show, the assumption that $\xi\gg 1$ is typically justified, as for quartic potential the amplitude of the curvature perturbation has the correct value only for large~$\xi$, given that the quartic self-coupling of the inflaton field is not very small, $\lambda\gtrsim \mathcal{O}(10^{-10})$. However, because scenarios where such small couplings are dynamically generated do exist (see e.g. \cite{Alanne:2016mpa}), in the following we will evaluate the potential numerically, without relying on analytical approximations which are shown here only for illustration.

Let us then discuss the case where the Jordan frame potential is quadratic in the field, i.e. $V(\phi)=m^2 \phi^2/2$. We will show that realizing successful inflation requires in this case $\xi\ll 1$, regardless of the choice of gravitational degrees of freedom. We see that in this limit the second term in Eq.~\eqref{chi_solution} becomes negligible and the metric theory asymptotes its Palatini counterpart. The Einstein frame potential then becomes in both cases
\bea
\label{U2}
U^{(2)}(\chi) &=& \frac{m^2}{2}\frac{\phi^2(\chi)}{\Omega^4(\phi(\chi))}= \frac{m^2M_{\rm P}^2}{2\xi}
\frac{{\rm sinh}^2\left(\sqrt{\xi}\chi/M_{\rm P}\right)}{\left(1+{\rm sinh}^2\left(\sqrt{\xi}\chi/M_{\rm P}\right)\right)^2} \\ \nonumber
&\approx& \frac{1}{2}m^2\chi^2 - \frac{5}{6}\xi\frac{m^2}{M_{\rm P}^2}\chi^4 + \mathcal{O}(\chi^6) ,
\eea
from which one can see that even though there is no plateau in this case, the non-minimal coupling still does have a flattening effect on the potential at large $\chi$ (see the left panel of Fig.~\ref{potential_plots}). However, we again emphasize that the above expressions are only for illustration and that in our numerical analysis we compute the inflationary observables using the exact potential in the Einstein frame. Also, note that in this paper $\phi=S$ in the case of quadratic potential, i.e. we do not study cases where the Higgs mass term would dominate. 

Finally, we make a remark about quantum corrections, which in the case of plateau potentials have been shown to be mostly insignificant during inflation (for recent works see e.g. \cite{Bilandzic:2007nb,Lerner:2009xg,Herranen:2016xsy,Fumagalli:2016sof,Markkanen:2017tun} and \cite{DeSimone:2008ei,Bezrukov:2009db,George:2013iia,George:2015nza,Bezrukov:2014ipa,Saltas:2015vsc,Bezrukov:2014bra,Bezrukov:2017dyv} for Higgs inflation specifically) but which might affect the potential in the regime where reheating occurs \cite{Bezrukov:2014bra,Bezrukov:2014ipa}. While such effects may be able to change our results to some extent, quantifying their exact effect is not only difficult but also certainly model-dependent. In this paper our aim is to study inflationary dynamics at classical level as accurately as possible without making an attempt to relate them with physics at lower energies. However, when it comes to the specific case where the SM Higgs drives inflation, we will assume that the threshold corrections to couplings at intermediate scales are large enough to support Higgs to act as the inflaton but small enough not to change the classical analysis to large extent. For detailed studies on these effects, see \cite{Bezrukov:2014ipa,Enckell:2016xse,Bezrukov:2017dyv,Enckell:2018kkc,Rasanen:2017ivk}. In case of other models, we neglect the possible effect of quantum corrections. This should not be a problem as our estimates for the reheating temperature and, consequently, inflationary observables are rather conservative, as we will discuss in more detail Section \ref{reheating}.


\subsection{Starobinsky inflation}
\label{starobinsky}

The case of Starobinsky inflation deserves special attention\footnote{For a historical review of Starobinsky inflation and non-minimally coupled models, see the Appendix of Ref. \cite{Hinshaw:2012aka} and references therein.}. For a general $f(R)$ theory, the gravitational part of the Jordan frame action is
\be 
\label{starobinsky_action}
	S_J = \frac12\int d^4x \sqrt{-g}\left( f(g_{\mu\nu},R_{\mu\nu}(\Gamma)) \right) \,,
\ee
which can be written dynamically equivalently as~\cite{Sotiriou:2008rp}
\be 
\label{starobinsky_action}
	S_J = \frac12\int d^4x \sqrt{-g}\left(f(z)+ f'(z)\left(g^{\mu\nu}R_{\mu\nu}-z\right)\right) \,,
\ee
where $z$ is an auxiliary field and we denote a derivative with respect to $z$ by prime. Variation with respect to $z$ leads to
\be
f''(\chi)\left(g^{\mu\nu}R_{\mu\nu}-z\right) = 0,
\ee
from which we see that for $f''(\chi)\neq 0$, the auxiliary field is $z=g^{\mu\nu}R_{\mu\nu}$.

Redefining the field by $\omega=f'(z)/M_{\rm P}$ and making again a Weyl transformation
\be
	g_{\mu\nu} \to \Omega^{2}g_{\mu\nu} = \frac{\omega}{M_{\rm P}^2} g_{\mu\nu} ,
\ee
gives the Einstein frame action
\be 
\label{starobinsky_action}
	S_E = \int d^4x \sqrt{-g}\left(\frac{1}{2}M_{\rm P}^2 R-\frac{3\kappa M_{\rm P}^2}{4\omega^2} g^{\mu\nu} \partial_{\mu}\omega\partial_{\nu}\omega - \frac{M_{\rm P}^3\left(\omega z-\displaystyle\frac{f(z)}{M_{\rm P}}\right)}{2\omega^2} \right) \,,
\ee
where again $\kappa=1$ in the metric case and $\kappa=0$ in the Palatini case. Because in the Palatini case the field $\omega$ is non-dynamical, we see that $f(R)$ gravity-driven inflation only has a metric counterpart\footnote{Even though in the Palatini case the field $\omega$ is non-dynamical and cannot act as an inflaton, it can take part in inflationary dynamics in the mixed case where $f(R)$ gravity is considered alongside a dynamical field driving inflation \cite{Enckell:2018hmo,Antoniadis:2018ywb,Antoniadis:2018yfq}.}. Therefore, concerning Starobinsky inflation, in the remaining of this paper we will consider only the case where $\kappa=1$. 

As in Section \ref{higgslike}, also in this case one can redefine the field by
\be
\frac{{\rm d}\chi}{{\rm d}\omega} = \sqrt{\frac{3}{2}}\frac{M_{\rm P}}{\omega} ,
\ee
so that the kinetic term becomes canonical and an action similar to \eqref{EframeS1} is retained. For Starobinsky inflation $f(R)=M_{\rm P}^2 R+\alpha R^2$, 
so in this case $z=R$ and $\omega/M_{\rm P}={\rm exp}(\sqrt{2/3}\chi/M_{\rm P})$. Thus the potential for the `scalaron' field $\chi$ becomes
\be
\label{UStarobinsky}
U(\chi) = \frac{M_{\rm P}^4}{8\alpha} \left(1-e^{-\sqrt{\frac{2}{3}}\frac{\chi}{M_{\rm P}}} \right)^2 ,
\ee
which closely resembles the potential for the metric case, Eq.~\eqref{chipotential1}, for $\chi\gg \sqrt{3/2}M_{\rm P}$ and with the identification $\alpha = \xi^2/2\lambda$. This is the origin of the famous $\xi$-attractor behavior \cite{Kaiser:2013sna,Kallosh:2013tua}, where independently of the original scalar potential in the Jordan frame the models asymptote to a universal attractor, the Starobinsky model. However, as was recently pointed out in \cite{Jarv:2017azx}, this behavior is not universal for non-minimally coupled models but depend on the choice of the gravitational degrees of freedom, as can be directly seen by comparing the Starobinsky potential \eqref{UStarobinsky} to the potential in the Palatini case, Eq.~\eqref{chipotential1}. Therefore, it should not come as a surprise that only the metric counterpart of Higgs-like inflation will be close to the Starobinsky model in the limit of large coupling.


\subsection{Inflationary observables}
\label{observables}

We will now move on to study the important inflationary observables. In slow-roll approximation the inflationary dynamics is characterized by the slow-roll parameters
\bea
\label{SRparameters1}
	\epsilon &\equiv& \frac{1}{2}M_{\rm P}^2 \left(\frac{U'}{U}\right)^2 \,, \quad
	\eta \equiv M_{\rm P}^2 \frac{U''}{U} \,, \quad \\ \nonumber
	\sigma_2 &\equiv& M_{\rm P}^4 \frac{U'}{U}\frac{U'''}{U} \,, \quad
	\hspace{3mm}\sigma_3 \equiv M_{\rm P}^6 \left(\frac{U'}{U}\right)^2\frac{U''''}{U} \,, \quad
\eea
where the prime denotes derivative with respect to $\chi$, and the total number of $e$-folds during inflation $N$ is given by
\be \label{Ndef}
	N = \frac{1}{M_{\rm P}^2} \int_{\chi_f}^{\chi_i} {\rm d}\chi \, U \left(\frac{{\rm d}U}{{\rm d} \chi}\right)^{-1}.
\ee
The field value at the end of inflation, $\chi_f$, is defined via $\epsilon(\chi_f)= 1$. The leading order expressions for the spectral index, its running and running of the running, and the tensor-to-scalar ratio are
\bea
\label{nsralpha}
n_s &\simeq& 1 - 6\epsilon + 2\eta\,, \\ \nonumber
\alpha_s &\simeq& -24\epsilon^2 + 16\epsilon\eta -2\sigma_2\,, \\ \nonumber
\beta_s &\simeq& -192\epsilon^3 + 192\epsilon^2\eta -32\epsilon\eta^2 - 24\epsilon\sigma_2 +2\eta\sigma_2 + 2\sigma_3\,, \\ \nonumber
r&\simeq& 16\epsilon\,,
\eea
respectively.

Let us first discuss the case where the Jordan frame potential is quartic in the field. 
Although the scalaron potential in the Starobinsky model, \eqref{UStarobinsky}, is very similar to the scalar field potential in the metric case, \eqref{chipotential1},
their predictions for the observables are slightly different in practice, and hence
here we also give the predictions for those in the Starobinsky case. 
We obtain
\bea 
\label{nsr1}
	n_s^{(4)} -1&\simeq& 
	\begin{cases} 
			-\displaystyle\frac{2}{N} + \frac{3}{2N^2}  & \quad \hspace{2mm} \mathrm{metric} , \vspace{2mm} \\
			-\displaystyle\frac{2}{N} - \frac{9}{2N^2}  & \quad \hspace{2mm} \mathrm{Starobinsky} , \vspace{2mm}\\ 
		-\displaystyle\frac{2}{N} - \frac{3}{8\xi N^2}  & \quad \hspace{2mm} \mathrm{Palatini},  \\
\end{cases}
\\ \notag \\
	\alpha_s^{(4)} &\simeq& 	
	\begin{cases}
		-\displaystyle\frac{2}{N^2}+\frac{9}{2N^3}  & \quad \mathrm{metric} ,\vspace{2mm} \\ 
			-\displaystyle\frac{2}{N^2}-\frac{21}{2N^3}  & \quad \mathrm{Starobinsky} ,\vspace{2mm}  \\
		-\displaystyle\frac{2}{N^2}-\frac{3}{4\xi N^3}  & \quad \mathrm{Palatini} , \\
	\end{cases}
	\\  \notag \\
	\beta_s^{(4)} &\simeq&
		\begin{cases}
		-\displaystyle\frac{4}{N^3}+\frac{33}{2N^4}  & \quad \mathrm{metric} ,  \vspace{2mm} \\
			-\displaystyle\frac{4}{N^3}-\frac{69}{2N^4}  & \quad \mathrm{Starobinsky} ,  \vspace{2mm} \\
		-\displaystyle\frac{4}{N^3}-\frac{9}{4\xi N^4}  & \quad \mathrm{Palatini} , \\
	\end{cases}
	\\  \notag \\
	\label{eq:r_N}
	r^{(4)} &\simeq& 
	\begin{cases}
		\displaystyle\frac{12}{N^2}-\frac{18}{N^3}  & \quad \mathrm{metric} ,  \vspace{2mm} \\
		\displaystyle\frac{12}{N^2}+\frac{18}{N^3}  & \quad \mathrm{Starobinsky} ,  \vspace{2mm} \\
		\displaystyle\frac{2}{\xi N^2}+\frac{1}{4\xi^2 N^3} & \quad \mathrm{Palatini} . \\
	\end{cases}
\eea
These approximate expressions were computed by neglecting the lower limit of integration and terms linear in the field in Eq.~\eqref{Ndef}. The results are in reasonable agreement with the results computed numerically from Eqs.~\eqref{chi1} and \eqref{Omega1} as explained in Section \ref{higgslike}, and Eq.~\eqref{UStarobinsky}. The values are only shown for illustration to elaborate differences between the different cases, and in Section \ref{results} we only show the results of numerical computation which are much more accurate.

The  curvature power spectrum is given by ~\cite{Lyth:1998xn,Aghanim:2018eyx}
\be
\label{cobe}
\mathcal{P}_{\zeta} = \frac{1}{24 \pi^2 M_{\rm P}^4} \frac{U(\chi_i)}{\epsilon(\chi_i)} ,
\ee
and the observed amplitude is $\mathcal{P}_{\zeta}=2.1\times 10^{-9}$ \cite{Akrami:2018odb}, which 
relates the non-minimal coupling to the number of required $e$-folds and the model parameters.
In the  case where the quartic self-coupling dominates the inflaton potential, we get
\be \label{xicondition1}
	\xi^{(4)} \simeq 
	\begin{cases}
		\sqrt{\displaystyle\frac{\lambda}{72\pi^2\mathcal{P}_{\zeta}}}N & \quad \mathrm{metric} , \vspace{3mm} \\
		\displaystyle\frac{\lambda N^2}{12\pi^2 \mathcal{P}_{\zeta}} & \quad \mathrm{Palatini} . \\
	\end{cases}
\ee
In the case of Starobinsky inflation, the correct amplitude for the curvature power spectrum is obtained for 
\be
\alpha\simeq \frac{N^2}{144\pi^2\mathcal{P}_{\zeta}} .
\ee

The most recent analysis of observations of the CMB made by the Planck satellite give (all at the $68\%$ confidence level) \cite{Akrami:2018odb}
\bea 
n_s &=& 0.9625\pm 0.0048, \\ \nonumber
\alpha_s &=& 0.002\pm 0.010, \\ \nonumber
\beta_s &=& 0.010\pm 0.013,
\label{Planck_const_run}
\eea
for the TT,TE,EE+lowE+lensing dataset including running of running of the spectral index. The joint analysis of data by Planck and BICEP2/Keck Array give (at the $95\%$ confidence level) \cite{Ade:2018gkx}
\be
r<0.06 .
\ee  
All quantities above are constrained at the pivot scale $k_*=0.05\, {\rm Mpc}^{-1}$. 
As mentioned in the introduction, although the current constraints on  the runnings $\alpha_s$ and $\beta_s$ are not  severe enough to test the inflationary models, 
they are expected to be improved much in future galaxy surveys and/or observations of the 21 cm line 
\cite{Kohri:2013mxa,Basse:2014qqa,Munoz:2016owz,Pourtsidou:2016ctq,Sekiguchi:2017cdy,Li:2018epc}. For example, in future observations of 21 cm fluctuations, $\alpha_s$ and  $\beta_s$ 
could be probed at the level of $ \alpha_s = {\cal O}(10^{-3})$ and $ \beta_s = {\cal O}(10^{-4})$, respectively \cite{Sekiguchi:2017cdy}.

In the metric case the predicted tensor-to-scalar ratio is well within reach of current or planned future experiments such as BICEP3 \cite{Wu:2016hul}, LiteBIRD~\cite{Matsumura:2013aja}, and the Simons Observatory \cite{Simons_Observatory}, but in the Palatini case the predicted tensor-to-scalar ratio is within reach of the current or near future experiments only if the non-minimal coupling takes a very small value $\xi\sim 1$, which requires $\lambda \lesssim 10^{-10}$. However, as in all cases the exact predictions for inflationary parameters depend not only on the values of these parameters but also on the total number of $e$-folds, we will postpone presenting the results until we have discussed reheating and how it affects the required number of $e$-folds. This will be done in Section \ref{reheating}, and the results are then shown in Section \ref{results}.

Let us then discuss the case where the Jordan frame potential is quadratic. In this case, it is difficult to find analytic estimates for the inflationary observables and we will perform a purely numerical evaluation instead. Similar to the quartic case, the measured amplitude of the curvature power spectrum fixes the mass parameter. One finds 
\be
m\simeq 5\times 10^{-6}M_{\rm P} ,
\ee
which shows why the SM Higgs with $m_{\rm H}=125$ GeV cannot act as an inflaton with a quadratic Jordan frame potential.

Before completing this section, let us briefly discuss models where the non-minimal coupling to gravity is absent. 
For the spectral index and tensor-to-scalar ratio, one obtains in this case the well-known results
\begin{eqnarray}
\label{nsr3}
n^{(4)}_{s;\xi=0}(\phi_i) = 1-\frac{3}{N+1} &\simeq& [0.941,  0.951], \hspace{1cm} r^{(4)}_{\xi=0}(\phi_i) = \frac{16}{N+1}\simeq  [0.26,  0.31], \nn \\ \nn
n^{(2)}_{s;\xi=0}(\phi_i) = 1-\frac{16}{8N+1} &\simeq& [0.960,  0.967], \hspace{1cm} r^{(2)}_{\xi=0}(\phi_i)=\frac{64}{8N+1} \simeq [0.13,  0.16],
\end{eqnarray}
for quartic and quadratic potentials, respectively. The numerical values apply for $N~=~[50,  60]$. As is evident, the results are now strongly disfavored by the Planck data. Together with the renormalizability requirements discussed in Section \ref{model}, this notion gives further motivation for introducing non-minimal couplings to gravity.


\section{Reheating}
\label{reheating}

As discussed above, in order to be able to distinguish different models from each other, one has to know their predictions for inflationary observables, such as $n_s$ and $r$, as accurately as possible. As shown above, the predictions depend on the required number of $e$-folds $N$, 
that is on the amount of expansion between the times when the pivot scale exited the horizon and the end of inflation.
How many $e$-folds are needed obviously depends on the expansion history after inflation, which in turn depends on the moment when the Universe was reheated, i.e. the reheating temperature. In the following, we will thus study reheating and the required number of $e$-folds in the above models case by case.


\subsection{Reheating in Higgs inflation}

Reheating in Higgs inflation has been studied exhaustively in the metric case in \cite{Bezrukov:2008ut,GarciaBellido:2008ab} (see also \cite{Ema:2016dny,Repond:2016sol}) but never before in the Palatini case. In the following, we will review the main results in the metric case and present the first calculations on reheating dynamics in the Palatini-Higgs inflation. 


\subsubsection{Metric case}

We begin by reviewing the main results in the metric case. In this case, inflation ends at $\chi_{\rm end}\simeq \sqrt{3/2} M_{\rm P} \log (1+2/\sqrt{3})$. The effective Einstein frame $\chi$ condensate then begins to oscillate with an initial field amplitude of $\chi  \lesssim \chi_{\rm end}$
as soon as the field becomes effectively massive, $U''/H^2\geq 1$. The potential is at this point effectively quadratic 
\be
U(\chi)\simeq \frac12\omega_{\rm H}^2\chi^2, 
\ee
with an effective mass $\omega_{\rm H}^2 = \lambda_{\rm H} M_{\rm P}^2/3\xi^2_{\rm H}$. 

The analysis presented in \cite{Bezrukov:2008ut,GarciaBellido:2008ab} shows that in this regime the Higgs condensate rapidly produces weak gauge bosons which subsequently decay into all other SM particles and reheat the Universe. The number of  $e$-folds is given by
\be
\label{efolds_higgs}
N = \ln\left(\left(\frac{\rho_{\rm RH}}{\rho_{\rm end}}\right)^{1/3}\left(\frac{g_0T_0^3}{g_*T_{\rm RH}^3} \right)^{1/3}H_k k^{-1} \right) \simeq 54.4 + \frac16\ln\left(\frac{r(N)}{10^{-3}}\right) + \frac13\ln\left(\frac{T_{\rm RH}}{10^{14}{\rm GeV}}\right) ,
\ee
where $\rho_{\rm RH}=(\pi^2/30) g_*T_{\rm RH}^4$, $\rho_{\rm end}\simeq z^2H_k^2M_{\rm P}^2$ with $z\equiv H_{\rm end}/H_k\simeq 0.54$, 
which characterizes how much $H$ changes between horizon exit of the pivot scale and the end of inflation.
In Fig.~\ref{Hratios}, we plot $z$ as a function of $\xi$ for the cases with quartic and quadratic Jordan frame potential both in the metric and Palatini cases. 
From the figure, we can approximate $z \simeq 0.54$ for the quartic metric case when the non-minimal coupling is large enough, $\xi\gtrsim 1$.
Also, $g_0 = 2+21/11\simeq 3.909$ is the entropic effective degrees of freedom at the present time.
For the Hubble parameter at the pivot scale, $k_*=0.05\, {\rm Mpc}^{-1}$, we use $H_k=7.84\times 10^{13}\sqrt{r/0.1}$ GeV. By using Eq.~\eqref{nsr1} for $r=r(N)$, the result~\eqref{efolds_higgs} can be expressed in terms of $T_{\rm RH}$ as
\be
\label{efolds2}
N = \frac13 W\left({\rm Exp}\left(169.0+{\rm ln\left(\frac{T_{\rm RH}}{10^{14}{\rm GeV}}\right)}\right)\right) ,
\ee
where $W$ is the principal branch of the Lambert W-function\footnote{The Lambert W-function, also called the omega or product log function, is the inverse function of\linebreak $f(x)=x e^x$. This is exactly what we get upon substituting Eq.~\eqref{nsr1} for $r=r(N)$ into Eq.~\eqref{efolds_higgs}.}. The reheating temperature is \cite{Bezrukov:2008ut}
\be
T_{\rm RH} = (3-15)\times 10^{13} {\rm GeV} ,
\ee
and therefore for the metric counterpart of Higgs inflation $N\simeq 54.2-54.8$. The result is in accord with the estimates given in \cite{Lerner:2011ge,Bezrukov:2011gp} when evaluated at the reference scale $k=0.002\, {\rm Mpc}^{-1}$ used in the above papers.


\subsubsection{Palatini case}
\label{Higgs_reheating_palatini}

So far, reheating dynamics has been studied in the case of Palatini inflation in \cite{Fu:2017iqg}, where the authors considered self-resonance of the inflaton field in a simple $\lambda\phi^4$ theory only, and in \cite{Almeida:2018oid}, where reheating in an $S$-inflation model was studied. In the following, we will follow Ref.~\cite{Almeida:2018oid} and present the first calculations on reheating dynamics in the Palatini-Higgs inflation, deriving expressions for the reheating temperature and the required number of $e$-folds.

\begin{figure}
\begin{center}
\includegraphics[width=.485\textwidth]{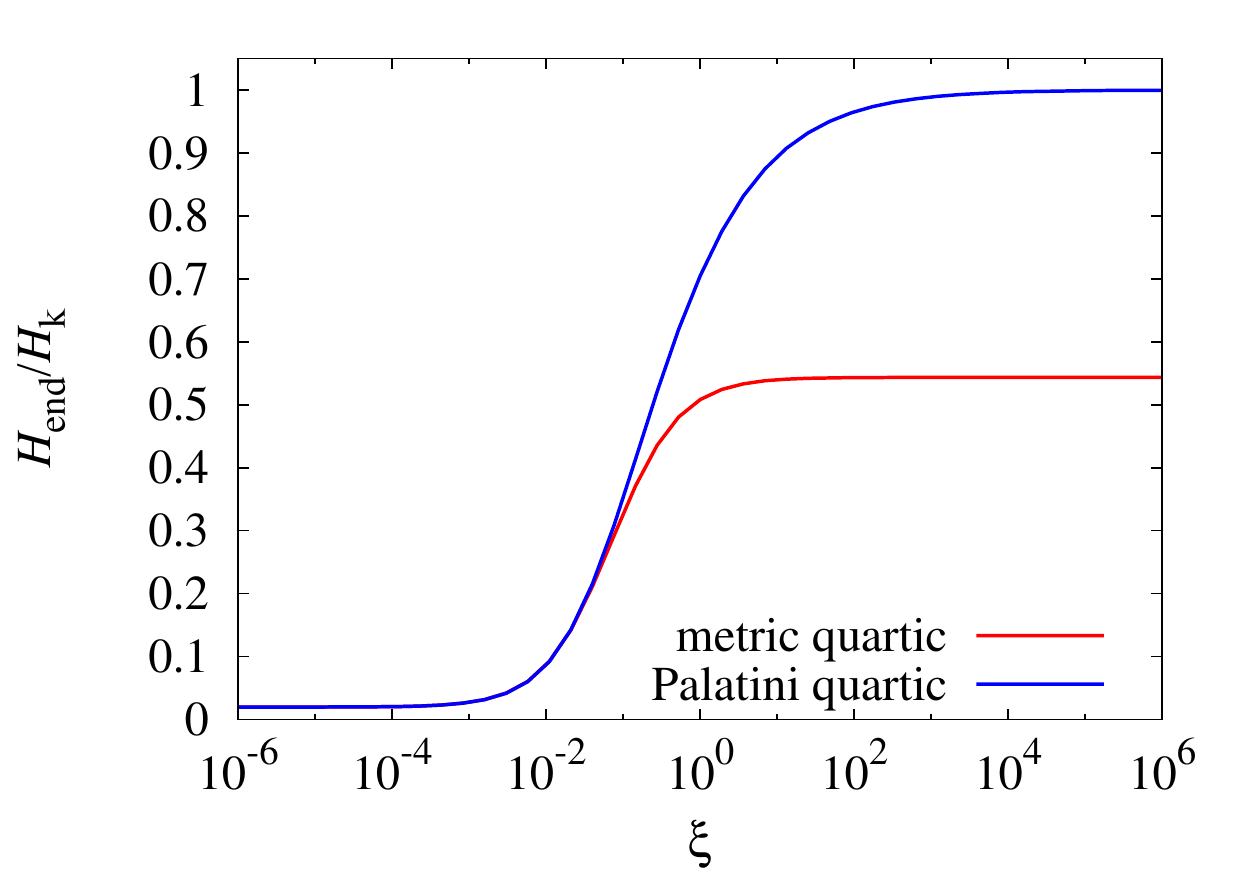}
\includegraphics[width=.485\textwidth]{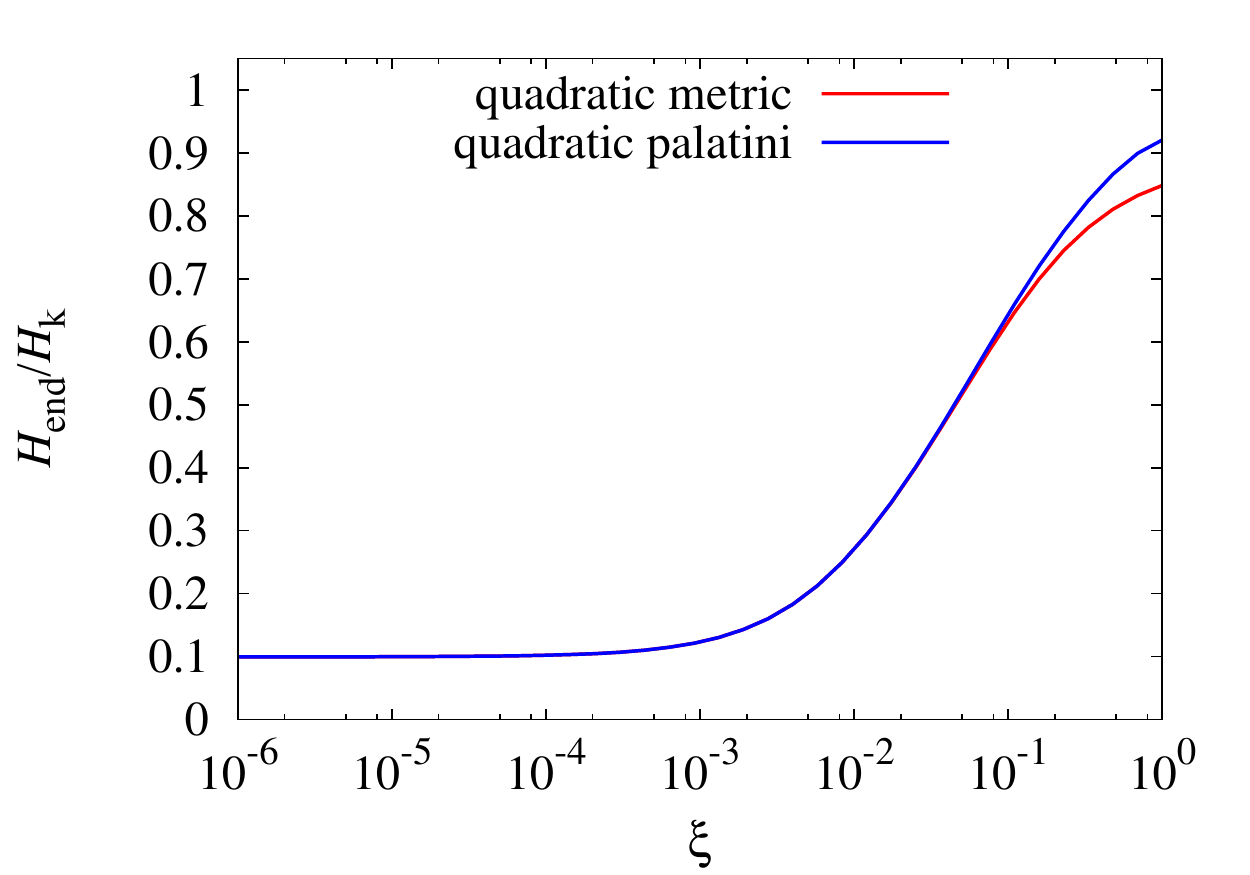}
\caption{Comparison of the Hubble parameter at the end of inflation and at the time when the pivot scale $k_*=0.05{\rm Mpc}^{-1}$ exited the horizon, defined as $H_{\rm end}/H_k=\sqrt{U(\chi_f)/U(\chi_i)}$. The left panel shows the ratio in case of quartic Jordan frame potential, $V(\phi)=\lambda\phi^4/4$, and the right panel in case of quadratic Jordan frame potential $V(\phi)= m^2\phi^2/2$. We have checked that at the limit $\xi\to 0$ the ratio $H_{\rm end}/H_k$ approaches the minimally coupled cases with quadratic and quartic potentials.}
\label{Hratios}
\end{center}
\end{figure}

In the Palatini case, inflation ends at $\chi_{\rm end}/M_{\rm P}=\arcsinh(4\sqrt{2\xi_{\rm H}})/(2\sqrt{\xi_{\rm H}})$. Thus, after the amplitude of inflaton oscillations has decreased down to $\chi_{\rm q}/M_{\rm P}\simeq 1/\sqrt{\xi}$, the inflaton potential can be approximated by
\be
U(\chi) \simeq \frac{\lambda_{\rm H}}{4}\chi^4.
\ee
The potential after inflation is almost exactly quartic\footnote{
By solving the Higgs' equation of motion numerically in the true $\tanh^4(\sqrt{\xi_{\rm H}}\xi/M_{\rm P})$ potential, 
we find that the quartic part of the potential is reached in less than one $e$-fold after inflation for all $\xi_{\rm H}$ under consideration.
}, so that the total number of required $e$-folds can be expressed independently of the reheating temperature as
\bea
\label{efolds3}
N &=& \ln\left(\left(\frac{\rho_{\rm RH}}{\rho_{\rm end}}\right)^{1/4}\left(\frac{g_0T_0^3}{g_*T_{\rm RH}^3} \right)^{1/3}H_k k^{-1} \right)  
\\ \nonumber
&\simeq& 55.1 + \frac14\ln\left(\frac{r(N,\xi_{\rm H})}{10^{-3}}\right) 
\eea
where we used $H_{\rm end}=H_k$ (see  Fig.~\ref{Hratios}). The total number of $e$-folds can then be expressed as
\be
\label{Higgs_Palatini_N}
N = \frac12 W\left({\rm Exp}\left(105.4-{\rm ln}\sqrt{\frac{\xi_{\rm H}}{10^8}}\right) \right) ,
\ee
where we again used Eq.~\eqref{nsr1} to exchange $r$ with $N$ and $\xi_{\rm H}$. 

As one can see, the result does not depend on the reheating temperature due to the fact that in this case the Higgs potential after inflation is almost exactly quartic, and as a result the Universe is effectively radiation-dominated from the end of inflation. The required number of $e$-folds still depends on the model parameters, namely $\xi_{\rm H}$, and in Section \ref{results} we will show the results for Palatini-Higgs inflation for $10^6\leq \xi \leq 10^{10}$. This corresponds to $10^{-4}\lesssim \lambda_{\rm H} \lesssim 1$, where the lower limit has been chosen to avoid fine-tuning in the SM beta functions. Therefore, for the Palatini counterpart of Higgs inflation, $N\simeq 49.2-51.5$.

Despite the fact that in this case the inflationary observables themselves are not sensitive to the reheating temperature, one can derive an estimate for $T_{\rm RH}$ in the Palatini-Higgs inflation. Taking $H^2\simeq \lambda_{\rm H}\chi^4/(12M_{\rm P}^2)$ after the quartic part of the Higgs potential has been reached and the field has started to oscillate about the minimum of its potential, and \cite{Ichikawa:2008ne,Kainulainen:2016vzv}
\be
\Gamma_{\chi\rightarrow W^+W^-} \simeq 5\times 10^{-4}g^4\lambda_{\rm H}^{-1/4}\chi ,
\ee
for the effective (semi-)perturbative decay rate of the Higgs into a pair of $W^\pm$ gauge bosons with $g$ the associated gauge coupling, we find that from the start of oscillations $\Gamma_{\chi\rightarrow W^+W^-}/H~>~1$ and the field decays immediately. The reheating temperature can then be estimated as
\be
\label{RH_temp_Higgs_Palatini}
T_{\rm RH}\simeq 2.5 \times 10^{14} ~{\rm GeV} \frac{N}{\sqrt{\xi_H}} \,,
\ee
where we used Eq.~\eqref{xicondition1} for $\lambda_{\rm H}/\xi_{\rm H}$ and assumed the usual number of SM degrees of freedom, $g_*=106.75$, as well as instant thermalization of SM particles after the decay of the Higgs condensate. 

Obviously, the above estimate neglects several phenomena known to play a role in preheating dynamics, such as backreaction of produced particles and the fact that in the beginning of post-inflationary era the potential is not exactly quartic. A detailed study on these effects will be presented elsewhere but we believe that for the purposes of the present paper the above estimates are sufficient, as in the Palatini case the Higgs potential indeed is quartic rather than quadratic from the end of inflation to a good accuracy, and hence the reheating temperature plays no significant role in determining the inflationary observables. Knowing the exact reheating temperature might, however, be important for other phenomena, such as post-inflationary phase transitions \cite{Enqvist:2014zqa,Cosme:2018wfh} or dark matter production, leptogenesis models \cite{Kusenko:2014lra}, primordial black hole formation \cite{Carr:2018nkm}, or models where something else than the inflaton-Higgs is responsible for producing the observed curvature perturbation at large scales, such as modulated reheating \cite{Dvali:2003em} or curvaton models \cite{Enqvist:2001zp,Lyth:2001nq,Moroi:2001ct}. It is in this spirit that we have presented our estimate for the reheating temperature, Eq.~\eqref{RH_temp_Higgs_Palatini}.

\subsection{Reheating in S-inflation}

Reheating in S-inflation has been studied in the metric case in \cite{Lerner:2011ge,Tenkanen:2016idg,Tenkanen:2016twd} and in \cite{Almeida:2018oid} in the Palatini case. Here we will review the main results obtained in these two cases.


\subsubsection{Quartic potential: metric case}

Again, after inflation the Einstein frame $\chi$ condensate begins to oscillate with an initial field value $\chi \simeq \chi_{\rm end}$.
The field oscillates first in a quadratic potential $U(\chi)=\omega_{\rm S}^2\chi^2/2$ with an effective mass $\omega_{\rm S}^2 = \lambda_{\rm S} M_{\rm P}^2/3\xi^2_{\rm S}$, until a transition into a quartic potential occurs at $\chi_{\rm q}/M_{\rm P}\simeq \sqrt{2/3}/\xi_{\rm S}$, unless the condensate decays before that.

The possible decay channels for an oscillating $S$ condensate are decay to SM particles and production of inflaton excitations \cite{Lerner:2011ge}. In the case the condensate decays while oscillating in the quadratic part of its potential, the number of inflationary $e$-folds is again given by Eq.~\eqref{efolds2}, where in this case $T_{\rm RH}$ is a free parameter which can be determined once the couplings of the $S$ field to the SM sector have been specified. However, one can derive an estimate for the smallest reheating temperature which is consistent with the assumption that the field decays while in quadratic potential, namely
\be
T_{\rm RH}^{\rm min} \simeq 4.5 \times 10^{13} \left(\frac{N}{60}\right)^{-1/2}\left( \frac{\xi_S}{10^4}\right)^{-1/2}{\rm GeV} .
\ee
The result means that in the case of $S$-inflation in the metric theory of gravity, Eq.~\eqref{efolds2} only applies down to this temperature. One can easily verify that the reheating temperature in the metric version of Higgs inflation satisfies this bound, i.e. $T_{\rm RH}^{\rm Higgs}>T_{\rm RH}^{\rm min}$. We note in passing that observationally there are no lower bounds on the reheating temperature besides the requirement that the SM radiation had to be in equilibrium by the time of Big Bang Nucleosynthesis (BBN), i.e.  $T_{\rm RH}>T_{\rm BBN}=\mathcal{O}(1)$ MeV.

On the other hand, if the $S$ condensate reaches the quartic part of its potential before decaying into SM particles, the result will not depend on the reheating temperature, similarly to the Palatini-Higgs case discussed in Section \ref{Higgs_reheating_palatini}. In that case, the number of inflationary $e$-folds is given by
\be
\label{efolds}
N = \ln\left(\left(\frac{\rho_{\rm RD}}{\rho_{\rm end}}\right)^{1/3}\left(\frac{\rho_{\rm RH}}{\rho_{\rm RD}}\right)^{1/4}\left(\frac{g_0T_0^3}{g_*T_{\rm RH}^3} \right)^{1/3}H_k k^{-1} \right) \simeq 57.3 + \frac16\ln\left(\frac{r(N,\xi_{\rm S})}{10^{-3}}\right) + \frac{1}{12}\ln\left(\frac{\lambda_{\rm S}}{\xi_{\rm S}^4}\right) ,
\ee
where the energy density at the time the scalar undergoes a transition to the quartic potential and the Universe enters into a radiation dominated era is $\rho_{\rm RD}=\lambda_{\rm S}M_{\rm P}^4/(9\xi_{\rm S}^4)$ \cite{Bezrukov:2008ut}, where $\lambda_{\rm S}/\xi_{\rm S}^2$ is again given by Eq.~\eqref{xicondition1}. The result can then be written as
\be
\label{efolds_Smetric}
N = \frac12 W\left({\rm Exp}\left(113.1-\frac13{\rm ln}\left(\frac{\xi_{\rm S}}{10^4}\right)\right)\right) ,
\ee
where we used Eqs.~\eqref{nsr1} and \eqref{xicondition1}. Here we take $1\leq \xi_{\rm S} \leq 10^{5}$ (corresponding roughly to $10^{-10}\lesssim \lambda_{\rm S}\lesssim 1$), so that for the metric counterpart of $S$-inflation with a quartic potential we find the maximum range $N\simeq 53.8-55.9$, which is a conservative result as well as the one obtained from Eq.~\eqref{efolds2}
 for the chosen range of $\xi_{\rm S}$. For the analysis on $\xi_{\rm S} < 1$, we refer to \cite{Almeida:2018oid}.


\subsubsection{Quartic potential: Palatini case}

In the Palatini version of $S$-inflation the reheating dynamics is exactly the same as in the Palatini counterpart of Higgs inflation, and hence the required number of $e$-folds is in this case given by 
Eq.~\eqref{Higgs_Palatini_N} with the substitution $\xi_{\rm H}\rightarrow \xi_{\rm S}$. Because now $\lambda_{\rm S}$ (and hence also $\xi_{\rm S}$) can be considerably smaller than $\lambda_{\rm H}$ 
($\xi_{\rm H}$)  without excessive fine-tuning, one can expect that in models where the quartic self-coupling is small the results will deviate from those of Palatini-Higgs inflation. 
Here we take $1\leq \xi_{\rm S} \leq 10^{10}$ (corresponding again to $10^{-10}\lesssim \lambda_{\rm S}\lesssim 1$), so that for the Palatini counterpart of $S$-inflation with a quartic potential $N\simeq 49.2-55.0$. For the analysis on $\xi_{\rm S} < 1$, see again Ref.~\cite{Almeida:2018oid}.

Finally, we note that even though the required number of $e$-folds does not in this case depend on the reheating temperature, the value of $T_{\rm RH}$ can again be determined once the couplings of the $S$ field to the SM sector have been specified, as in \cite{Almeida:2018oid}.


\subsubsection{Quadratic potential}

Finally, we will study reheating in the case where the potential of the Jordan frame $S$ field is quadratic, rather than quartic, during inflation. After inflation, the Einstein frame potential for the field $\chi$ is given by Eq.~\eqref{U2}, and the field quickly reaches the purely quadratic part of its potential so that $\Omega^2(S)\rightarrow 1$ and the Einstein and Jordan frames become equivalent. The field starts to oscillate and can then decay into SM particles but not fragment directly into inflaton particles, as this channel is kinematically blocked \cite{Kainulainen:2016vzv}.

The number of $e$-folds is again given by an expression similar to Eq.~\eqref{efolds_higgs}, namely
\be
N = \ln\left(\left(\frac{\rho_{\rm RH}}{\rho_{\rm end}}\right)^{1/3}\left(\frac{g_0T_0^3}{g_*T_{\rm RH}^3} \right)^{1/3}H_k k^{-1} \right) \simeq 54.4 + \frac16\ln\left(\frac{r(N,\xi_{\rm S})}{0.01}\right) + \frac13\ln\left(\frac{z^{-2}T_{\rm RH}}{ 10^{14}{\rm GeV}}\right) ,
\ee
where due to large variation in $z\equiv H_{\rm end}/H_k\simeq 0.12$ (see Fig. \ref{Hratios}), we have written the result explicitly in terms of it. We will evaluate $z$ numerically for each set of parameters. 
The reheating temperature we again take to be a free parameter, and allow for a large range of values between $10^6~{\rm GeV}$ and $10^{16}~{\rm GeV}$. 
In Section \ref{results}, we present the results for $n_s, r, \alpha_s$ and $\beta_s$ with $\xi_{\rm S}$ and $T_R$ being varied.
For $\xi_{\rm S} = 10^{-3}$,  we numerically find $N\simeq 50.0 - 57.8 $ with the above mentioned range of $T_R$.

Because in this case the non-minimal coupling to gravity is required to be very small, $\xi_{\rm S}\simeq \mathcal{O}(10^{-3})$, basically the only difference in reheating between the metric and Palatini counterparts of gravity is in the value of $z$. This difference is typically very small, which makes it very difficult to distinguish between different theories of gravity through observations in the case where the inflaton potential is quadratic during inflation. We will discuss this further in Section \ref{results}.


\subsection{Reheating in Starobinsky inflation}

Reheating in Starobinsky inflation has been studied exhaustively in \cite{Gorbunov:2010bn} and here we again review the main results. As noted in Section \ref{starobinsky}, there is no Palatini counterpart for Starobinsky inflation, and hence we will study only the metric case.

In the case of Starobinsky inflation the scalaron couplings to all fields are suppressed by $M_{\rm P}$, and reheating mainly occurs via decay of the scalaron into the SM Higgs bosons, which then annihilate into other SM particles thus reheating the Universe. The reheating temperature is \cite{Gorbunov:2010bn}
\be
\label{RH_Starobinsky}
T_{\rm RH} = 3\times 10^{9} {\rm GeV} .
\ee
Because the potential in the Starobinsky case is similar to that of metric Higgs inflation also during reheating, the total number of required $e$-folds is again given by Eq.~\eqref{efolds2}. For the above result for $T_{\rm RH}$ we find $N=51.2$, in accord with the results presented in \cite{Bezrukov:2011gp} when $N$ is evaluated at the reference scale $k=0.002\, {\rm Mpc}^{-1}$  used in that paper.


\section{Results and discussion}
\label{results}

\begin{figure}
\begin{center}
\includegraphics[width=1.0\textwidth]{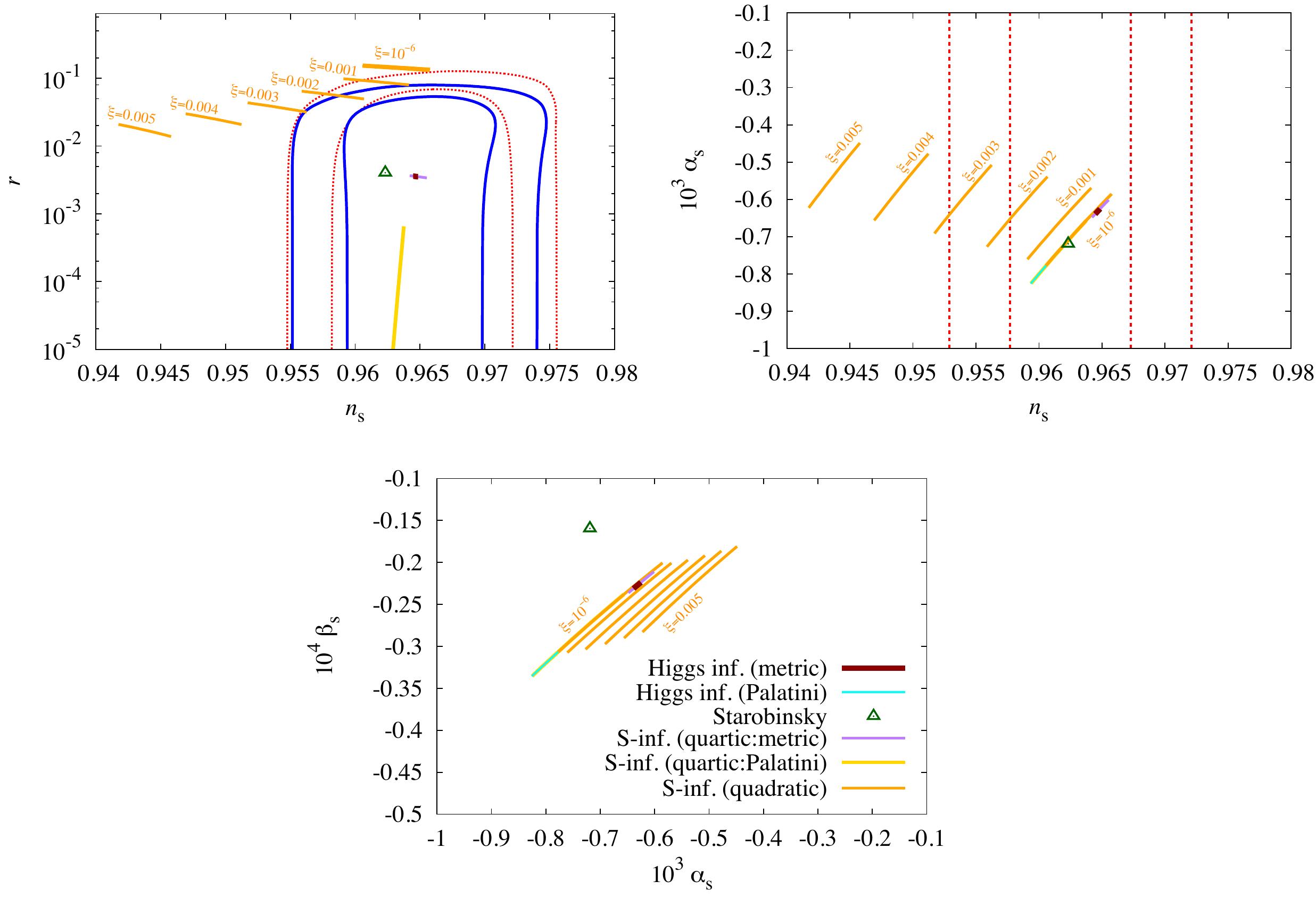}
\caption{Predictions for inflationary observables. {\it Upper left panel}: Spectral index $n_s$ vs. tensor-to-scalar ratio $r$. {\it Upper right panel}: Spectral index $n_s$ vs. its running $\alpha_s$. {\it Lower  panel}: Running of the spectral index $\alpha_s$  vs. running of the running $\beta_s$. The color associated to each model is indicated in the legend in the lower  panel. For comparison,
we also plot in the upper left panel the 1$\sigma$ and 2$\sigma$ constraints on $n_s$ and $r$ from Planck TT, TE, EE+lowE+lensing (red dotted) and TT, TE, EE+lowE+lensing+BK14 (blue solid) \cite{Akrami:2018odb}. In the upper right panel, we show 1$\sigma$ and 2$\sigma$ constraints on $n_s$ denoted in Eq.~\eqref{Planck_const_run} where we have quoted from the analysis allowing $\alpha_s$ and $\beta_s$ varied.  
In the case of $\alpha_s$ and $\beta_s$, the observational limits are outside the boundaries of the plots. For the quadratic S-inflation case, we show the prediction for several values of $\xi$, as given in the Figure. Also in other cases $\xi$ is varied as discussed in the main text.}
\label{results_plot}
\end{center}
\end{figure}

Predictions for the inflationary observables $n_s, r, \alpha_s$, and $\beta_s$ of the models discussed in the previous sections are shown in Fig.~\ref{results_plot}. These plots are our main results. We reiterate that we have evaluated all observables numerically from Eqs.~\eqref{SRparameters1}--\eqref{nsralpha} and \eqref{cobe}, without relying on the analytical estimates given in Section \ref{observables}. In the Palatini version of Higgs inflation the prediction for tensor-to-scalar ratio is very small, $r \ll 10^{-5}$ (see Eq.~\eqref{eq:r_N}), and therefore not shown in Fig.~\ref{results_plot}. For comparison with data, we have plotted the constraints on $n_s$ and $r$ from Planck \cite{Akrami:2018odb}. In the case of the running parameters $\alpha_s$ and $\beta_s$, the observational limits are outside the boundaries of the plots. Although the current constraints on them are weak, they can be 
much improved by using, e.g. future observations of galaxy surveys and/or the 21 cm line \cite{Kohri:2013mxa,Basse:2014qqa,Munoz:2016owz,Pourtsidou:2016ctq,Sekiguchi:2017cdy,Li:2018epc}. As already mentioned, $\alpha_s$ and $\beta_s$ 
could be probed at the level of $ \alpha_s = {\cal O}(10^{-3})$ and $ \beta_s = {\cal O}(10^{-4})$, respectively, for example, in future observations of 21 cm fluctuations \cite{Sekiguchi:2017cdy}.

In the case of quadratic S-inflation, we only show the metric case since the predictions of the Palatini case are practically indistinguishable from those of the metric case. Since the observables are in this case very sensitive to the value of $\xi$, we show the predictions for several values of it. We also note that the small disrepancy between the results reported here and those presented in Ref.~\cite{Tenkanen:2017jih} by one of the present authors is due to an error in the numerical computation of slow-roll parameters in \cite{Tenkanen:2017jih}.

The results show that already by measuring a non-zero tensor-to-scalar ratio $r$ most scenarios discussed in this paper can be distinguished from each other, i.e. positive observation of $r$ would rule out most models. A notable exception to this is the case where the inflaton has a quartic potential in the Jordan frame and gravity is of metric type, as then both the SM Higgs (depicted by the brown line) and an inflaton field other than the SM Higgs (depicted by the purple line) can predict $r\simeq 0.0035 , n_s\simeq 0.965$.
As can be seen in Fig.~\ref{results_plot}, it may not be possible to break this degeneracy by measuring running or running of the running of the spectral index either, as the predictions would still be too similar to each other (unless quantum corrections strongly affect the potentials). However, by specifying a model and computing the reheating temperature -- and hence the required number of $e$-folds -- accurately, it may still be possible to distinguish these two cases from each other. As noted above, observations are also unlikely to separate the metric counterpart from the Palatini one in the case where the inflaton potential in the Jordan frame is quadratic in the field. These cases are, however, easily distinguishable from models where the potential is not quadratic.

Another case where observations are unlikely to be able to distinguish between models of the same type is the one where the inflaton has a quartic potential in the Jordan frame, gravity is of Palatini type, and the SM Higgs has a very small quartic self-coupling, $\lambda_{\rm H}\lesssim 10^{-5}$, at the inflationary scale. However, as in this case the prediction for tensor-to-scalar ratio is typically very small, $r\lesssim 10^{-5}$, this scenario is less interesting from the observational point of view. As discussed in Section \ref{introduction}, on-going or near-future CMB B-mode polarization experiments such as BICEP3 \cite{Wu:2016hul}, LiteBIRD~\cite{Matsumura:2013aja} and the Simons Observatory \cite{Simons_Observatory} would be pushing the limit on tensor-to-scalar ratio down to $r\lesssim 0.001$, or aiming to detect $r$ above this limit. This would probe Palatini models with quartic potentials only when the inflaton has a small coupling to gravity, $\xi<1$, which corresponds to $\lambda\lesssim 10^{-10}$. For detailed analysis of such scenario, see \cite{Almeida:2018oid}.

Similar analysis as conducted here could also be done, for example, for all the variants of Higgs inflation. Indeed, as models where the SM Higgs assumes the role of inflaton have been shown to exhibit rich and versatile phenomenology (see e.g. \cite{Bezrukov:2007ep,Bauer:2008zj,Germani:2010gm,Nakayama:2010kt,Kamada:2010qe,Kamada:2012se,Jinno:2017lun,Enckell:2018kkc,Rasanen:2018ihz}), it would be interesting to see if measuring the main inflationary observables could distinguish between these variants.


\section{Conclusions}
\label{conclusions}

In this paper we have made a detailed comparison between different non-minimal models for cosmic inflation. We studied models where the inflaton is either coupled non-minimally to gravity via $\xi \phi^2 R$, or it is the dynamical scalar degree of freedom of the Starobinsky model where the gravity sector is enlarged by an $R^2$ term. We considered the original Higgs inflation, Starobinsky inflation, and two different versions of a scenario where the inflaton is a scalar field other than the SM Higgs, and discussed to what extent these models can be distinguished from each other by measuring the main inflationary observables. Assuming slow-roll, we calculated predictions for the spectral index, tensor-to-scalar ratio, and running and running of the running of the spectral index. We considered non-minimal couplings in both metric and Palatini theories of gravity, and discussed if inflationary observables can also provide ways to distinguish between different theories of gravity in the context of the models under consideration. We followed the evolution of the fields through reheating to compute how many $e$-folds between horizon exit of a reference scale, where observables are measured, and the end of inflation each model predicts, which is necessary to compute the observables accurately.

We found that the current and near-future CMB B-mode polarization experiments are generically able to distinguish between models of the same type, although a notable exception to this is the case where the inflaton has a quartic potential in the Jordan frame and gravity is of metric type, as then both the SM Higgs and a model where inflation is driven by a field other than the SM Higgs can predict too similar values for inflationary observables, at least when the reheating temperature (and hence the number of $e$-folds) is not accurately known.

It would be interesting to conduct a similar analysis also for other models where the inflaton is coupled non-minimally to gravity and/or the gravity sector is otherwise extended. This type of models typically have a flattening effect on the inflaton potential and therefore predict smaller values for the tensor-to-scalar ratio than models without such couplings, and are thus capable of resurrecting such models. Determining the inflaton potential and, ultimately, the high energy theory responsible for inflation therefore requires detailed calculations of dynamics of the inflaton field both during and after inflation. We hope that our work paves the way for achieving this goal.


\section*{Acknowledgements}
We thank K. Aoki, N. Arkani-Hamed, I. Bah, J.P. Beltr\'{a}n Almeida, K. Berghaus, N. Bernal, P. Carrilho, T. Clifton, R. Jinno, D.E. Kaplan, T. Koivisto, S. Mukohyama, D. Mulryne, M. Oliosi, J. Rubio, S. Rusak, J. Sakstein, J. Yokoyama, L.-P. Wahlman, and Y. Watanabe for correspondence and discussions. T. Tenkanen acknowledges Saga University, Yukawa Institute for Theoretical Physics, and RESCEU for hospitality. 
The work of T.~Takahashi was supported by JSPS KAKENHI Grant Number 15K05084,  17H01131 and MEXT KAKENHI Grant Number 15H05888.
T.~Tenkanen is funded by the Simons foundation, the U.K. Science and Technology Facilities Council grant ST/J001546/1, and the Magnus Ehrnrooth foundation.


\bibliography{HI_variants}

\providecommand{\href}[2]{#2}\begingroup\raggedright\begin{thebibliography}{100}

\bibitem{Starobinsky:1980te}
A.~A. Starobinsky, {\it {A New Type of Isotropic Cosmological Models Without
  Singularity}},  {\em Phys. Lett.} {\bf B91} (1980) 99--102. [771(1980)].

\bibitem{Sato:1980yn}
K.~Sato, {\it {First Order Phase Transition of a Vacuum and Expansion of the
  Universe}},  {\em Mon.Not.Roy.Astron.Soc.} {\bf 195} (1981) 467--479.

\bibitem{Guth:1980zm}
A.~H. Guth, {\it {The Inflationary Universe: A Possible Solution to the Horizon
  and Flatness Problems}},  {\em Phys.Rev.} {\bf D23} (1981) 347--356.

\bibitem{Linde:1981mu}
A.~D. Linde, {\it {A New Inflationary Universe Scenario: A Possible Solution of
  the Horizon, Flatness, Homogeneity, Isotropy and Primordial Monopole
  Problems}},  {\em Phys.Lett.} {\bf B108} (1982) 389--393.

\bibitem{Albrecht:1982wi}
A.~Albrecht and P.~J. Steinhardt, {\it {Cosmology for Grand Unified Theories
  with Radiatively Induced Symmetry Breaking}},  {\em Phys.Rev.Lett.} {\bf 48}
  (1982) 1220--1223.

\bibitem{Linde:1983gd}
A.~D. Linde, {\it {Chaotic Inflation}},  {\em Phys.Lett.} {\bf B129} (1983)
  177--181.

\bibitem{Lyth:1998xn}
D.~H. Lyth and A.~Riotto, {\it {Particle physics models of inflation and the
  cosmological density perturbation}},  {\em Phys. Rept.} {\bf 314} (1999)
  1--146, [\href{http://arxiv.org/abs/hep-ph/9807278}{{\tt hep-ph/9807278}}].

\bibitem{Mazumdar:2010sa}
A.~Mazumdar and J.~Rocher, {\it {Particle physics models of inflation and
  curvaton scenarios}},  {\em Phys. Rept.} {\bf 497} (2011) 85--215,
  [\href{http://arxiv.org/abs/1001.0993}{{\tt arXiv:1001.0993}}].

\bibitem{Martin:2013tda}
J.~Martin, C.~Ringeval, and V.~Vennin, {\it {Encyclopaedia Inflationaris}},
  {\em Phys. Dark Univ.} {\bf 5-6} (2014) 75--235,
  [\href{http://arxiv.org/abs/1303.3787}{{\tt arXiv:1303.3787}}].

\bibitem{Patrignani:2016xqp}
{\bf Particle Data Group} Collaboration, C.~Patrignani et~al., {\it {Review of
  Particle Physics}},  {\em Chin. Phys.} {\bf C40} (2016), no.~10 100001.

\bibitem{Akrami:2018odb}
{\bf Planck} Collaboration, Y.~Akrami et~al., {\it {Planck 2018 results. X.
  Constraints on inflation}},  \href{http://arxiv.org/abs/1807.06211}{{\tt
  arXiv:1807.06211}}.

\bibitem{Aghanim:2018eyx}
{\bf Planck} Collaboration, N.~Aghanim et~al., {\it {Planck 2018 results. VI.
  Cosmological parameters}},  \href{http://arxiv.org/abs/1807.06209}{{\tt
  arXiv:1807.06209}}.

\bibitem{Ade:2018gkx}
{\bf BICEP2, Keck Array} Collaboration, P.~A.~R. Ade et~al., {\it {BICEP2 /
  Keck Array x: Constraints on Primordial Gravitational Waves using Planck,
  WMAP, and New BICEP2/Keck Observations through the 2015 Season}},  {\em
  Submitted to: Phys. Rev. Lett.} (2018)
  [\href{http://arxiv.org/abs/1810.05216}{{\tt arXiv:1810.05216}}].

\bibitem{Wu:2016hul}
W.~L.~K. Wu et~al., {\it {Initial Performance of BICEP3: A Degree Angular Scale
  95 GHz Band Polarimeter}},  {\em J. Low. Temp. Phys.} {\bf 184} (2016),
  no.~3-4 765--771, [\href{http://arxiv.org/abs/1601.00125}{{\tt
  arXiv:1601.00125}}].

\bibitem{Matsumura:2013aja}
T.~Matsumura et~al., {\it {Mission design of LiteBIRD}},
  \href{http://arxiv.org/abs/1311.2847}{{\tt arXiv:1311.2847}}. [J. Low. Temp.
  Phys.176,733(2014)].

\bibitem{Simons_Observatory}
{\bf Simons Observatory} Collaboration, P.~Ade et~al., {\it {The Simons
  Observatory: Science goals and forecasts}},
  \href{http://arxiv.org/abs/1808.07445}{{\tt arXiv:1808.07445}}.

\bibitem{Kohri:2013mxa}
K.~Kohri, Y.~Oyama, T.~Sekiguchi, and T.~Takahashi, {\it {Precise Measurements
  of Primordial Power Spectrum with 21 cm Fluctuations}},  {\em JCAP} {\bf
  1310} (2013) 065, [\href{http://arxiv.org/abs/1303.1688}{{\tt
  arXiv:1303.1688}}].

\bibitem{Basse:2014qqa}
T.~Basse, J.~Hamann, S.~Hannestad, and Y.~Y.~Y. Wong, {\it {Getting leverage on
  inflation with a large photometric redshift survey}},  {\em JCAP} {\bf 1506}
  (2015), no.~06 042, [\href{http://arxiv.org/abs/1409.3469}{{\tt
  arXiv:1409.3469}}].

\bibitem{Munoz:2016owz}
J.~B. Muñoz, E.~D. Kovetz, A.~Raccanelli, M.~Kamionkowski, and J.~Silk, {\it
  {Towards a measurement of the spectral runnings}},  {\em JCAP} {\bf 1705}
  (2017) 032, [\href{http://arxiv.org/abs/1611.05883}{{\tt arXiv:1611.05883}}].

\bibitem{Pourtsidou:2016ctq}
A.~Pourtsidou, {\it {Synergistic tests of inflation}},
  \href{http://arxiv.org/abs/1612.05138}{{\tt arXiv:1612.05138}}.

\bibitem{Sekiguchi:2017cdy}
T.~Sekiguchi, T.~Takahashi, H.~Tashiro, and S.~Yokoyama, {\it {21 cm Angular
  Power Spectrum from Minihalos as a Probe of Primordial Spectral Runnings}},
  {\em JCAP} {\bf 1802} (2018), no.~02 053,
  [\href{http://arxiv.org/abs/1705.00405}{{\tt arXiv:1705.00405}}].

\bibitem{Li:2018epc}
X.~Li, N.~Weaverdyck, S.~Adhikari, D.~Huterer, J.~Muir, and H.-Y. Wu, {\it {The
  Quest for the Inflationary Spectral Runnings in the Presence of Systematic
  Errors}},  {\em Astrophys. J.} {\bf 862} (2018), no.~2 137,
  [\href{http://arxiv.org/abs/1806.02515}{{\tt arXiv:1806.02515}}].

\bibitem{Salopek:1988qh}
D.~S. Salopek, J.~R. Bond, and J.~M. Bardeen, {\it {Designing Density
  Fluctuation Spectra in Inflation}},  {\em Phys. Rev.} {\bf D40} (1989) 1753.

\bibitem{Bezrukov:2007ep}
F.~L. Bezrukov and M.~Shaposhnikov, {\it {The Standard Model Higgs boson as the
  inflaton}},  {\em Phys. Lett.} {\bf B659} (2008) 703--706,
  [\href{http://arxiv.org/abs/0710.3755}{{\tt arXiv:0710.3755}}].

\bibitem{Rubio:2018ogq}
J.~Rubio, {\it {Higgs inflation}},  \href{http://arxiv.org/abs/1807.02376}{{\tt
  arXiv:1807.02376}}.

\bibitem{Spokoiny:1984bd}
B.~L. Spokoiny, {\it {Inflation and generation of perturbations in broken
  symmetric theory of gravity}},  {\em Phys. Lett.} {\bf 147B} (1984) 39--43.

\bibitem{Futamase:1987ua}
T.~Futamase and K.-i. Maeda, {\it {Chaotic Inflationary Scenario in Models
  Having Nonminimal Coupling With Curvature}},  {\em Phys. Rev.} {\bf D39}
  (1989) 399--404.

\bibitem{Fakir:1990eg}
R.~Fakir and W.~G. Unruh, {\it {Improvement on cosmological chaotic inflation
  through nonminimal coupling}},  {\em Phys. Rev.} {\bf D41} (1990) 1783--1791.

\bibitem{Amendola:1990nn}
L.~Amendola, M.~Litterio, and F.~Occhionero, {\it {The Phase space view of
  inflation. 1: The nonminimally coupled scalar field}},  {\em Int. J. Mod.
  Phys.} {\bf A5} (1990) 3861--3886.

\bibitem{Kaiser:1994vs}
D.~I. Kaiser, {\it {Primordial spectral indices from generalized Einstein
  theories}},  {\em Phys. Rev.} {\bf D52} (1995) 4295--4306,
  [\href{http://arxiv.org/abs/astro-ph/9408044}{{\tt astro-ph/9408044}}].

\bibitem{Komatsu:1999mt}
E.~Komatsu and T.~Futamase, {\it {Complete constraints on a nonminimally
  coupled chaotic inflationary scenario from the cosmic microwave background}},
   {\em Phys. Rev.} {\bf D59} (1999) 064029,
  [\href{http://arxiv.org/abs/astro-ph/9901127}{{\tt astro-ph/9901127}}].

\bibitem{Bauer:2008zj}
F.~Bauer and D.~A. Demir, {\it {Inflation with Non-Minimal Coupling: Metric
  versus Palatini Formulations}},  {\em Phys. Lett.} {\bf B665} (2008)
  222--226, [\href{http://arxiv.org/abs/0803.2664}{{\tt arXiv:0803.2664}}].

\bibitem{Park:2008hz}
S.~C. Park and S.~Yamaguchi, {\it {Inflation by non-minimal coupling}},  {\em
  JCAP} {\bf 0808} (2008) 009, [\href{http://arxiv.org/abs/0801.1722}{{\tt
  arXiv:0801.1722}}].

\bibitem{Lerner:2009xg}
R.~N. Lerner and J.~McDonald, {\it {Gauge singlet scalar as inflaton and
  thermal relic dark matter}},  {\em Phys. Rev.} {\bf D80} (2009) 123507,
  [\href{http://arxiv.org/abs/0909.0520}{{\tt arXiv:0909.0520}}].

\bibitem{Linde:2011nh}
A.~Linde, M.~Noorbala, and A.~Westphal, {\it {Observational consequences of
  chaotic inflation with nonminimal coupling to gravity}},  {\em JCAP} {\bf
  1103} (2011) 013, [\href{http://arxiv.org/abs/1101.2652}{{\tt
  arXiv:1101.2652}}].

\bibitem{Kaiser:2013sna}
D.~I. Kaiser and E.~I. Sfakianakis, {\it {Multifield Inflation after Planck:
  The Case for Nonminimal Couplings}},  {\em Phys. Rev. Lett.} {\bf 112}
  (2014), no.~1 011302, [\href{http://arxiv.org/abs/1304.0363}{{\tt
  arXiv:1304.0363}}].

\bibitem{Kallosh:2013maa}
R.~Kallosh and A.~Linde, {\it {Non-minimal Inflationary Attractors}},  {\em
  JCAP} {\bf 1310} (2013) 033, [\href{http://arxiv.org/abs/1307.7938}{{\tt
  arXiv:1307.7938}}].

\bibitem{Kallosh:2013tua}
R.~Kallosh, A.~Linde, and D.~Roest, {\it {Universal Attractor for Inflation at
  Strong Coupling}},  {\em Phys. Rev. Lett.} {\bf 112} (2014), no.~1 011303,
  [\href{http://arxiv.org/abs/1310.3950}{{\tt arXiv:1310.3950}}].

\bibitem{Chiba:2014sva}
T.~Chiba and K.~Kohri, {\it {Consistency Relations for Large Field Inflation:
  Non-minimal Coupling}},  {\em PTEP} {\bf 2015} (2015), no.~2 023E01,
  [\href{http://arxiv.org/abs/1411.7104}{{\tt arXiv:1411.7104}}].

\bibitem{Boubekeur:2015xza}
L.~Boubekeur, E.~Giusarma, O.~Mena, and H.~Ramírez, {\it {Does Current Data
  Prefer a Non-minimally Coupled Inflaton?}},  {\em Phys. Rev.} {\bf D91}
  (2015) 103004, [\href{http://arxiv.org/abs/1502.05193}{{\tt
  arXiv:1502.05193}}].

\bibitem{Pieroni:2015cma}
M.~Pieroni, {\it {$\beta$-function formalism for inflationary models with a non
  minimal coupling with gravity}},  {\em JCAP} {\bf 1602} (2016), no.~02 012,
  [\href{http://arxiv.org/abs/1510.03691}{{\tt arXiv:1510.03691}}].

\bibitem{Salvio:2017xul}
A.~Salvio, {\it {Inflationary Perturbations in No-Scale Theories}},  {\em Eur.
  Phys. J.} {\bf C77} (2017), no.~4 267,
  [\href{http://arxiv.org/abs/1703.08012}{{\tt arXiv:1703.08012}}].

\bibitem{Odintsov:2018qyy}
S.~D. Odintsov and V.~K. Oikonomou, {\it {Attractor cosmology from nonminimally
  coupled gravity}},  {\em Phys. Rev.} {\bf D97} (2018), no.~6 064005,
  [\href{http://arxiv.org/abs/1802.06486}{{\tt arXiv:1802.06486}}].

\bibitem{Almeida:2018pir}
J.~P. Beltrán~Almeida and N.~Bernal, {\it {Nonminimally coupled pseudoscalar
  inflaton}},  {\em Phys. Rev.} {\bf D98} (2018), no.~8 083519,
  [\href{http://arxiv.org/abs/1803.09743}{{\tt arXiv:1803.09743}}].

\bibitem{Ferreira:2018nav}
R.~Z. Ferreira, A.~Notari, and G.~Simeon, {\it {Natural Inflation with a
  periodic non-minimal coupling}},  {\em JCAP} {\bf 1811} (2018), no.~11 021,
  [\href{http://arxiv.org/abs/1806.05511}{{\tt arXiv:1806.05511}}].

\bibitem{Germani:2010gm}
C.~Germani and A.~Kehagias, {\it {New Model of Inflation with Non-minimal
  Derivative Coupling of Standard Model Higgs Boson to Gravity}},  {\em Phys.
  Rev. Lett.} {\bf 105} (2010) 011302,
  [\href{http://arxiv.org/abs/1003.2635}{{\tt arXiv:1003.2635}}].

\bibitem{Nakayama:2010kt}
K.~Nakayama and F.~Takahashi, {\it {Running Kinetic Inflation}},  {\em JCAP}
  {\bf 1011} (2010) 009, [\href{http://arxiv.org/abs/1008.2956}{{\tt
  arXiv:1008.2956}}].

\bibitem{Kamada:2010qe}
K.~Kamada, T.~Kobayashi, M.~Yamaguchi, and J.~Yokoyama, {\it {Higgs
  G-inflation}},  {\em Phys. Rev.} {\bf D83} (2011) 083515,
  [\href{http://arxiv.org/abs/1012.4238}{{\tt arXiv:1012.4238}}].

\bibitem{Kamada:2012se}
K.~Kamada, T.~Kobayashi, T.~Takahashi, M.~Yamaguchi, and J.~Yokoyama, {\it
  {Generalized Higgs inflation}},  {\em Phys. Rev.} {\bf D86} (2012) 023504,
  [\href{http://arxiv.org/abs/1203.4059}{{\tt arXiv:1203.4059}}].

\bibitem{Jinno:2017lun}
R.~Jinno, K.~Kaneta, and K.-y. Oda, {\it {Hill-climbing Higgs inflation}},
  {\em Phys. Rev.} {\bf D97} (2018), no.~2 023523,
  [\href{http://arxiv.org/abs/1705.03696}{{\tt arXiv:1705.03696}}].

\bibitem{Enckell:2018kkc}
V.-M. Enckell, K.~Enqvist, S.~Rasanen, and E.~Tomberg, {\it {Higgs inflation at
  the hilltop}},  {\em JCAP} {\bf 1806} (2018), no.~06 005,
  [\href{http://arxiv.org/abs/1802.09299}{{\tt arXiv:1802.09299}}].

\bibitem{Rasanen:2018ihz}
S.~Rasanen, {\it {Higgs inflation in the Palatini formulation with kinetic
  terms for the metric}},  \href{http://arxiv.org/abs/1811.09514}{{\tt
  arXiv:1811.09514}}.

\bibitem{Lerner:2011ge}
R.~N. Lerner and J.~McDonald, {\it {Distinguishing Higgs inflation and its
  variants}},  {\em Phys. Rev.} {\bf D83} (2011) 123522,
  [\href{http://arxiv.org/abs/1104.2468}{{\tt arXiv:1104.2468}}].

\bibitem{Bezrukov:2011gp}
F.~L. Bezrukov and D.~S. Gorbunov, {\it {Distinguishing between R$^2$-inflation
  and Higgs-inflation}},  {\em Phys. Lett.} {\bf B713} (2012) 365--368,
  [\href{http://arxiv.org/abs/1111.4397}{{\tt arXiv:1111.4397}}].

\bibitem{Sotiriou:2008rp}
T.~P. Sotiriou and V.~Faraoni, {\it {f(R) Theories Of Gravity}},  {\em Rev.
  Mod. Phys.} {\bf 82} (2010) 451--497,
  [\href{http://arxiv.org/abs/0805.1726}{{\tt arXiv:0805.1726}}].

\bibitem{Bauer:2010jg}
F.~Bauer and D.~A. Demir, {\it {Higgs-Palatini Inflation and Unitarity}},  {\em
  Phys. Lett.} {\bf B698} (2011) 425--429,
  [\href{http://arxiv.org/abs/1012.2900}{{\tt arXiv:1012.2900}}].

\bibitem{Tamanini:2010uq}
N.~Tamanini and C.~R. Contaldi, {\it {Inflationary Perturbations in Palatini
  Generalised Gravity}},  {\em Phys. Rev.} {\bf D83} (2011) 044018,
  [\href{http://arxiv.org/abs/1010.0689}{{\tt arXiv:1010.0689}}].

\bibitem{Rasanen:2017ivk}
S.~Rasanen and P.~Wahlman, {\it {Higgs inflation with loop corrections in the
  Palatini formulation}},  {\em JCAP} {\bf 1711} (2017), no.~11 047,
  [\href{http://arxiv.org/abs/1709.07853}{{\tt arXiv:1709.07853}}].

\bibitem{Tenkanen:2017jih}
T.~Tenkanen, {\it {Resurrecting Quadratic Inflation with a non-minimal coupling
  to gravity}},  {\em JCAP} {\bf 1712} (2017), no.~12 001,
  [\href{http://arxiv.org/abs/1710.02758}{{\tt arXiv:1710.02758}}].

\bibitem{Racioppi:2017spw}
A.~Racioppi, {\it {Coleman-Weinberg linear inflation: metric vs. Palatini
  formulation}},  {\em JCAP} {\bf 1712} (2017), no.~12 041,
  [\href{http://arxiv.org/abs/1710.04853}{{\tt arXiv:1710.04853}}].

\bibitem{Markkanen:2017tun}
T.~Markkanen, T.~Tenkanen, V.~Vaskonen, and H.~Veermäe, {\it {Quantum
  corrections to quartic inflation with a non-minimal coupling: metric vs.
  Palatini}},  {\em JCAP} {\bf 1803} (2018), no.~03 029,
  [\href{http://arxiv.org/abs/1712.04874}{{\tt arXiv:1712.04874}}].

\bibitem{Jarv:2017azx}
L.~Järv, A.~Racioppi, and T.~Tenkanen, {\it {Palatini side of inflationary
  attractors}},  {\em Phys. Rev.} {\bf D97} (2018), no.~8 083513,
  [\href{http://arxiv.org/abs/1712.08471}{{\tt arXiv:1712.08471}}].

\bibitem{Racioppi:2018zoy}
A.~Racioppi, {\it {New universal attractor in nonminimally coupled gravity:
  Linear inflation}},  {\em Phys. Rev.} {\bf D97} (2018), no.~12 123514,
  [\href{http://arxiv.org/abs/1801.08810}{{\tt arXiv:1801.08810}}].

\bibitem{Carrilho:2018ffi}
P.~Carrilho, D.~Mulryne, J.~Ronayne, and T.~Tenkanen, {\it {Attractor Behaviour
  in Multifield Inflation}},  {\em JCAP} {\bf 1806} (2018), no.~06 032,
  [\href{http://arxiv.org/abs/1804.10489}{{\tt arXiv:1804.10489}}].

\bibitem{Enckell:2018hmo}
V.-M. Enckell, K.~Enqvist, S.~Rasanen, and L.-P. Wahlman, {\it {Inflation with
  $R^2$ term in the Palatini formalism}},
  \href{http://arxiv.org/abs/1810.05536}{{\tt arXiv:1810.05536}}.

\bibitem{Antoniadis:2018ywb}
I.~Antoniadis, A.~Karam, A.~Lykkas, and K.~Tamvakis, {\it {Palatini inflation
  in models with an $R^2$ term}},  {\em JCAP} {\bf 1811} (2018), no.~11 028,
  [\href{http://arxiv.org/abs/1810.10418}{{\tt arXiv:1810.10418}}].

\bibitem{Rasanen:2018fom}
S.~Rasanen and E.~Tomberg, {\it {Planck scale black hole dark matter from Higgs
  inflation}},  \href{http://arxiv.org/abs/1810.12608}{{\tt arXiv:1810.12608}}.
  [JCAP1901,038(2019)].

\bibitem{Kannike:2018zwn}
K.~Kannike, A.~Kubarski, L.~Marzola, and A.~Racioppi, {\it {A minimal model of
  inflation and dark radiation}},  \href{http://arxiv.org/abs/1810.12689}{{\tt
  arXiv:1810.12689}}.

\bibitem{Almeida:2018oid}
J.~P.~B. Almeida, N.~Bernal, J.~Rubio, and T.~Tenkanen, {\it {Hidden Inflaton
  Dark Matter}},  {\em JCAP} {\bf 1903} (2019) 012,
  [\href{http://arxiv.org/abs/1811.09640}{{\tt arXiv:1811.09640}}].

\bibitem{Antoniadis:2018yfq}
I.~Antoniadis, A.~Karam, A.~Lykkas, T.~Pappas, and K.~Tamvakis, {\it {Rescuing
  Quartic and Natural Inflation in the Palatini Formalism}},
  \href{http://arxiv.org/abs/1812.00847}{{\tt arXiv:1812.00847}}.

\bibitem{Azri:2017uor}
H.~Azri and D.~Demir, {\it {Affine Inflation}},  {\em Phys. Rev.} {\bf D95}
  (2017), no.~12 124007, [\href{http://arxiv.org/abs/1705.05822}{{\tt
  arXiv:1705.05822}}].

\bibitem{Azri:2018gsz}
H.~Azri, {\it {Are there really conformal frames? Uniqueness of affine
  inflation}},  {\em Int. J. Mod. Phys.} {\bf D27} (2018), no.~09 1830006,
  [\href{http://arxiv.org/abs/1802.01247}{{\tt arXiv:1802.01247}}].

\bibitem{Shimada:2018lnm}
K.~Shimada, K.~Aoki, and K.-i. Maeda, {\it {Metric-affine Gravity and
  Inflation}},  \href{http://arxiv.org/abs/1812.03420}{{\tt arXiv:1812.03420}}.

\bibitem{Birrell:1982ix}
N.~D. Birrell and P.~C.~W. Davies, {\em {Quantum Fields in Curved Space}}.
\newblock Cambridge Monographs on Mathematical Physics. Cambridge Univ. Press,
  Cambridge, UK, 1984.

\bibitem{Atkins:2012yn}
M.~Atkins and X.~Calmet, {\it {Bounds on the Nonminimal Coupling of the Higgs
  Boson to Gravity}},  {\em Phys. Rev. Lett.} {\bf 110} (2013), no.~5 051301,
  [\href{http://arxiv.org/abs/1211.0281}{{\tt arXiv:1211.0281}}].

\bibitem{LanahanTremblay:2007sg}
N.~Lanahan-Tremblay and V.~Faraoni, {\it {The Cauchy problem of f(R) gravity}},
   {\em Class. Quant. Grav.} {\bf 24} (2007) 5667--5680,
  [\href{http://arxiv.org/abs/0709.4414}{{\tt arXiv:0709.4414}}].

\bibitem{Aoki:2018lwx}
K.~Aoki and K.~Shimada, {\it {Galileon and generalized Galileon with projective
  invariance in metric-affine formalism}},  {\em Phys. Rev.} {\bf D98} (2018),
  no.~4 044038, [\href{http://arxiv.org/abs/1806.02589}{{\tt
  arXiv:1806.02589}}].

\bibitem{Capozziello:2015lza}
S.~Capozziello, T.~Harko, T.~S. Koivisto, F.~S.~N. Lobo, and G.~J. Olmo, {\it
  {Hybrid metric-Palatini gravity}},  {\em Universe} {\bf 1} (2015), no.~2
  199--238, [\href{http://arxiv.org/abs/1508.04641}{{\tt arXiv:1508.04641}}].

\bibitem{Koivisto:2005yc}
T.~Koivisto and H.~Kurki-Suonio, {\it {Cosmological perturbations in the
  palatini formulation of modified gravity}},  {\em Class. Quant. Grav.} {\bf
  23} (2006) 2355--2369, [\href{http://arxiv.org/abs/astro-ph/0509422}{{\tt
  astro-ph/0509422}}].

\bibitem{Salvio:2015kka}
A.~Salvio and A.~Mazumdar, {\it {Classical and Quantum Initial Conditions for
  Higgs Inflation}},  {\em Phys. Lett.} {\bf B750} (2015) 194--200,
  [\href{http://arxiv.org/abs/1506.07520}{{\tt arXiv:1506.07520}}].

\bibitem{Calmet:2016fsr}
X.~Calmet and I.~Kuntz, {\it {Higgs Starobinsky Inflation}},  {\em Eur. Phys.
  J.} {\bf C76} (2016), no.~5 289, [\href{http://arxiv.org/abs/1605.02236}{{\tt
  arXiv:1605.02236}}].

\bibitem{Wang:2017fuy}
Y.-C. Wang and T.~Wang, {\it {Primordial perturbations generated by Higgs field
  and $R^2$ operator}},  {\em Phys. Rev.} {\bf D96} (2017), no.~12 123506,
  [\href{http://arxiv.org/abs/1701.06636}{{\tt arXiv:1701.06636}}].

\bibitem{Ema:2017rqn}
Y.~Ema, {\it {Higgs Scalaron Mixed Inflation}},  {\em Phys. Lett.} {\bf B770}
  (2017) 403--411, [\href{http://arxiv.org/abs/1701.07665}{{\tt
  arXiv:1701.07665}}].

\bibitem{He:2018gyf}
M.~He, A.~A. Starobinsky, and J.~Yokoyama, {\it {Inflation in the mixed
  Higgs-$R^2$ model}},  {\em JCAP} {\bf 1805} (2018), no.~05 064,
  [\href{http://arxiv.org/abs/1804.00409}{{\tt arXiv:1804.00409}}].

\bibitem{Ghilencea:2018rqg}
D.~M. Ghilencea, {\it {Two-loop corrections to Starobinsky-Higgs inflation}},
  {\em Phys. Rev.} {\bf D98} (2018), no.~10 103524,
  [\href{http://arxiv.org/abs/1807.06900}{{\tt arXiv:1807.06900}}].

\bibitem{Gundhi:2018wyz}
A.~Gundhi and C.~F. Steinwachs, {\it {Scalaron-Higgs inflation}},
  \href{http://arxiv.org/abs/1810.10546}{{\tt arXiv:1810.10546}}.

\bibitem{Karam:2018mft}
A.~Karam, T.~Pappas, and K.~Tamvakis, {\it {Nonminimal Coleman--Weinberg
  Inflation with an $R^2$ term}},  {\em JCAP} {\bf 1902} (2019) 006,
  [\href{http://arxiv.org/abs/1810.12884}{{\tt arXiv:1810.12884}}].

\bibitem{Enckell:2018uic}
V.-M. Enckell, K.~Enqvist, S.~Rasanen, and L.-P. Wahlman, {\it {Higgs-$R^2$
  inflation -- full slow-roll study at tree-level}},
  \href{http://arxiv.org/abs/1812.08754}{{\tt arXiv:1812.08754}}.

\bibitem{Schutz:2013fua}
K.~Schutz, E.~I. Sfakianakis, and D.~I. Kaiser, {\it {Multifield Inflation
  after Planck: Isocurvature Modes from Nonminimal Couplings}},  {\em Phys.
  Rev.} {\bf D89} (2014), no.~6 064044,
  [\href{http://arxiv.org/abs/1310.8285}{{\tt arXiv:1310.8285}}].

\bibitem{Kallosh:2013daa}
R.~Kallosh and A.~Linde, {\it {Multi-field Conformal Cosmological Attractors}},
   {\em JCAP} {\bf 1312} (2013) 006,
  [\href{http://arxiv.org/abs/1309.2015}{{\tt arXiv:1309.2015}}].

\bibitem{GarciaBellido:2008ab}
J.~Garcia-Bellido, D.~G. Figueroa, and J.~Rubio, {\it {Preheating in the
  Standard Model with the Higgs-Inflaton coupled to gravity}},  {\em Phys.
  Rev.} {\bf D79} (2009) 063531, [\href{http://arxiv.org/abs/0812.4624}{{\tt
  arXiv:0812.4624}}].

\bibitem{Alanne:2016mpa}
T.~Alanne, F.~Sannino, T.~Tenkanen, and K.~Tuominen, {\it {Inflation and
  pseudo-Goldstone Higgs boson}},  {\em Phys. Rev.} {\bf D95} (2017), no.~3
  035004, [\href{http://arxiv.org/abs/1611.04932}{{\tt arXiv:1611.04932}}].

\bibitem{Bilandzic:2007nb}
A.~Bilandzic and T.~Prokopec, {\it {Quantum radiative corrections to slow-roll
  inflation}},  {\em Phys. Rev.} {\bf D76} (2007) 103507,
  [\href{http://arxiv.org/abs/0704.1905}{{\tt arXiv:0704.1905}}].

\bibitem{Herranen:2016xsy}
M.~Herranen, A.~Hohenegger, A.~Osland, and A.~Tranberg, {\it {Quantum
  corrections to inflation: the importance of RG-running and choosing the
  optimal RG-scale}},  {\em Phys. Rev.} {\bf D95} (2017), no.~2 023525,
  [\href{http://arxiv.org/abs/1608.08906}{{\tt arXiv:1608.08906}}].

\bibitem{Fumagalli:2016sof}
J.~Fumagalli, {\it {Renormalization Group independence of Cosmological
  Attractors}},  {\em Phys. Lett.} {\bf B769} (2017) 451--459,
  [\href{http://arxiv.org/abs/1611.04997}{{\tt arXiv:1611.04997}}].

\bibitem{DeSimone:2008ei}
A.~De~Simone, M.~P. Hertzberg, and F.~Wilczek, {\it {Running Inflation in the
  Standard Model}},  {\em Phys. Lett.} {\bf B678} (2009) 1--8,
  [\href{http://arxiv.org/abs/0812.4946}{{\tt arXiv:0812.4946}}].

\bibitem{Bezrukov:2009db}
F.~Bezrukov and M.~Shaposhnikov, {\it {Standard Model Higgs boson mass from
  inflation: Two loop analysis}},  {\em JHEP} {\bf 07} (2009) 089,
  [\href{http://arxiv.org/abs/0904.1537}{{\tt arXiv:0904.1537}}].

\bibitem{George:2013iia}
D.~P. George, S.~Mooij, and M.~Postma, {\it {Quantum corrections in Higgs
  inflation: the real scalar case}},  {\em JCAP} {\bf 1402} (2014) 024,
  [\href{http://arxiv.org/abs/1310.2157}{{\tt arXiv:1310.2157}}].

\bibitem{George:2015nza}
D.~P. George, S.~Mooij, and M.~Postma, {\it {Quantum corrections in Higgs
  inflation: the Standard Model case}},  {\em JCAP} {\bf 1604} (2016), no.~04
  006, [\href{http://arxiv.org/abs/1508.04660}{{\tt arXiv:1508.04660}}].

\bibitem{Bezrukov:2014ipa}
F.~Bezrukov, J.~Rubio, and M.~Shaposhnikov, {\it {Living beyond the edge: Higgs
  inflation and vacuum metastability}},  {\em Phys. Rev.} {\bf D92} (2015),
  no.~8 083512, [\href{http://arxiv.org/abs/1412.3811}{{\tt arXiv:1412.3811}}].

\bibitem{Saltas:2015vsc}
I.~D. Saltas, {\it {Higgs inflation and quantum gravity: An exact
  renormalisation group approach}},  {\em JCAP} {\bf 1602} (2016) 048,
  [\href{http://arxiv.org/abs/1512.06134}{{\tt arXiv:1512.06134}}].

\bibitem{Bezrukov:2014bra}
F.~Bezrukov and M.~Shaposhnikov, {\it {Higgs inflation at the critical point}},
   {\em Phys. Lett.} {\bf B734} (2014) 249--254,
  [\href{http://arxiv.org/abs/1403.6078}{{\tt arXiv:1403.6078}}].

\bibitem{Bezrukov:2017dyv}
F.~Bezrukov, M.~Pauly, and J.~Rubio, {\it {On the robustness of the primordial
  power spectrum in renormalized Higgs inflation}},  {\em JCAP} {\bf 1802}
  (2018), no.~02 040, [\href{http://arxiv.org/abs/1706.05007}{{\tt
  arXiv:1706.05007}}].

\bibitem{Enckell:2016xse}
V.-M. Enckell, K.~Enqvist, and S.~Nurmi, {\it {Observational signatures of
  Higgs inflation}},  {\em JCAP} {\bf 1607} (2016), no.~07 047,
  [\href{http://arxiv.org/abs/1603.07572}{{\tt arXiv:1603.07572}}].

\bibitem{Hinshaw:2012aka}
{\bf WMAP} Collaboration, G.~Hinshaw et~al., {\it {Nine-Year Wilkinson
  Microwave Anisotropy Probe (WMAP) Observations: Cosmological Parameter
  Results}},  {\em Astrophys. J. Suppl.} {\bf 208} (2013) 19,
  [\href{http://arxiv.org/abs/1212.5226}{{\tt arXiv:1212.5226}}].

\bibitem{Bezrukov:2008ut}
F.~Bezrukov, D.~Gorbunov, and M.~Shaposhnikov, {\it {On initial conditions for
  the Hot Big Bang}},  {\em JCAP} {\bf 0906} (2009) 029,
  [\href{http://arxiv.org/abs/0812.3622}{{\tt arXiv:0812.3622}}].

\bibitem{Ema:2016dny}
Y.~Ema, R.~Jinno, K.~Mukaida, and K.~Nakayama, {\it {Violent Preheating in
  Inflation with Nonminimal Coupling}},  {\em JCAP} {\bf 1702} (2017), no.~02
  045, [\href{http://arxiv.org/abs/1609.05209}{{\tt arXiv:1609.05209}}].

\bibitem{Repond:2016sol}
J.~Repond and J.~Rubio, {\it {Combined Preheating on the lattice with
  applications to Higgs inflation}},  {\em JCAP} {\bf 1607} (2016), no.~07 043,
  [\href{http://arxiv.org/abs/1604.08238}{{\tt arXiv:1604.08238}}].

\bibitem{Fu:2017iqg}
C.~Fu, P.~Wu, and H.~Yu, {\it {Inflationary dynamics and preheating of the
  nonminimally coupled inflaton field in the metric and Palatini formalisms}},
  {\em Phys. Rev.} {\bf D96} (2017), no.~10 103542,
  [\href{http://arxiv.org/abs/1801.04089}{{\tt arXiv:1801.04089}}].

\bibitem{Ichikawa:2008ne}
K.~Ichikawa, T.~Suyama, T.~Takahashi, and M.~Yamaguchi, {\it {Primordial
  Curvature Fluctuation and Its Non-Gaussianity in Models with Modulated
  Reheating}},  {\em Phys. Rev.} {\bf D78} (2008) 063545,
  [\href{http://arxiv.org/abs/0807.3988}{{\tt arXiv:0807.3988}}].

\bibitem{Kainulainen:2016vzv}
K.~Kainulainen, S.~Nurmi, T.~Tenkanen, K.~Tuominen, and V.~Vaskonen, {\it
  {Isocurvature Constraints on Portal Couplings}},  {\em JCAP} {\bf 1606}
  (2016), no.~06 022, [\href{http://arxiv.org/abs/1601.07733}{{\tt
  arXiv:1601.07733}}].

\bibitem{Enqvist:2014zqa}
K.~Enqvist, S.~Nurmi, T.~Tenkanen, and K.~Tuominen, {\it {Standard Model with a
  real singlet scalar and inflation}},  {\em JCAP} {\bf 1408} (2014) 035,
  [\href{http://arxiv.org/abs/1407.0659}{{\tt arXiv:1407.0659}}].

\bibitem{Cosme:2018wfh}
C.~Cosme, J.~G. Rosa, and O.~Bertolami, {\it {Can dark matter drive electroweak
  symmetry breaking?}},  \href{http://arxiv.org/abs/1811.08908}{{\tt
  arXiv:1811.08908}}.

\bibitem{Kusenko:2014lra}
A.~Kusenko, L.~Pearce, and L.~Yang, {\it {Postinflationary Higgs relaxation and
  the origin of matter-antimatter asymmetry}},  {\em Phys. Rev. Lett.} {\bf
  114} (2015), no.~6 061302, [\href{http://arxiv.org/abs/1410.0722}{{\tt
  arXiv:1410.0722}}].

\bibitem{Carr:2018nkm}
B.~Carr, K.~Dimopoulos, C.~Owen, and T.~Tenkanen, {\it {Primordial Black Hole
  Formation During Slow Reheating After Inflation}},  {\em Phys. Rev.} {\bf
  D97} (2018), no.~12 123535, [\href{http://arxiv.org/abs/1804.08639}{{\tt
  arXiv:1804.08639}}].

\bibitem{Dvali:2003em}
G.~Dvali, A.~Gruzinov, and M.~Zaldarriaga, {\it {A new mechanism for generating
  density perturbations from inflation}},  {\em Phys. Rev.} {\bf D69} (2004)
  023505, [\href{http://arxiv.org/abs/astro-ph/0303591}{{\tt
  astro-ph/0303591}}].

\bibitem{Enqvist:2001zp}
K.~Enqvist and M.~S. Sloth, {\it {Adiabatic CMB perturbations in pre - big bang
  string cosmology}},  {\em Nucl. Phys.} {\bf B626} (2002) 395--409,
  [\href{http://arxiv.org/abs/hep-ph/0109214}{{\tt hep-ph/0109214}}].

\bibitem{Lyth:2001nq}
D.~H. Lyth and D.~Wands, {\it {Generating the curvature perturbation without an
  inflaton}},  {\em Phys. Lett.} {\bf B524} (2002) 5--14,
  [\href{http://arxiv.org/abs/hep-ph/0110002}{{\tt hep-ph/0110002}}].

\bibitem{Moroi:2001ct}
T.~Moroi and T.~Takahashi, {\it {Effects of cosmological moduli fields on
  cosmic microwave background}},  {\em Phys. Lett.} {\bf B522} (2001) 215--221,
  [\href{http://arxiv.org/abs/hep-ph/0110096}{{\tt hep-ph/0110096}}]. [Erratum:
  Phys. Lett.B539,303(2002)].

\bibitem{Tenkanen:2016idg}
T.~Tenkanen, K.~Tuominen, and V.~Vaskonen, {\it {A Strong Electroweak Phase
  Transition from the Inflaton Field}},  {\em JCAP} {\bf 1609} (2016), no.~09
  037, [\href{http://arxiv.org/abs/1606.06063}{{\tt arXiv:1606.06063}}].

\bibitem{Tenkanen:2016twd}
T.~Tenkanen, {\it {Feebly Interacting Dark Matter Particle as the Inflaton}},
  {\em JHEP} {\bf 09} (2016) 049, [\href{http://arxiv.org/abs/1607.01379}{{\tt
  arXiv:1607.01379}}].

\bibitem{Gorbunov:2010bn}
D.~S. Gorbunov and A.~G. Panin, {\it {Scalaron the mighty: producing dark
  matter and baryon asymmetry at reheating}},  {\em Phys. Lett.} {\bf B700}
  (2011) 157--162, [\href{http://arxiv.org/abs/1009.2448}{{\tt
  arXiv:1009.2448}}].

\end{thebibliography}\endgroup


\end{document}